\documentclass[aps,prc,reprint,amsmath,amssymb,superscriptaddress,showpacs]{revtex4-2}

\usepackage{newtxtext,newtxmath}
\usepackage{graphicx}% Include figure files
\usepackage{dcolumn}% Align table columns on decimal point
\usepackage{bm}% bold math
\usepackage[]{hyperref}
\usepackage{etoolbox}

\hypersetup{
    colorlinks=true, 
    linkcolor=blue, 
    filecolor=blue, 
    urlcolor=blue, 
    citecolor=blue,
}
\pretocmd{\ref}{\bfseries}{}{}
    
\newcommand{\energy}[1]{$\sqrt{s_{\text{NN}}}$ = #1 TeV}
\newcommand{\two}[2]{$#1_{\text{#2}}$}
\newcommand{\mtwo}[2]{#1_{\text{#2}}}
\newcommand{\fq}[1]{$F_{\text{#1}} \text{(M)}$}

\begin{document}

\preprint{APS/123-QED}

\title{Topological analysis of scale-invariant spatial fluctuations in heavy-ion collisions at ultra-relativistic energies}
\author{Salman Khurshid Malik}
     \email{salmankm@keemail.me}
\author{Ramni Gupta}
    \email{ramni.gupta@cern.ch}
\author{Fakhar Ul Haider}
\author{Balwan Singh}

\affiliation{Department of Physics, University of Jammu, J\&K, India, 180006}
\date{\today}

\begin{abstract}
The QGP-to-hadronic matter phase transition and QCD critical point in heavy-ion collisions can be identified by studying spatial fluctuations among final-state particles using intermittency analysis.  First CMC-based intermittency analysis in the two-dimensional angular ($\eta$, $\varphi$) phase space, using EPOS as the background model is presented. Critical fluctuation signals are extremely weak, constituting only a few percent of the total event sample and are severely diluted by the overwhelming non-critical background, rendering traditional intermittency analyses insufficient for reliable signal extraction. To extract the weak critical signal, we employ a two-stage topological machine learning framework combining Topological Data Analysis (TDA) with deep learning. In the first stage, particle events are represented as two-dimensional point clouds and a Delaunay-based sub-level set filtration is constructed to extract Betti curves as multiscale topological invariants, corrected for multiplicity bias via azimuthal randomisation and classified by two complementary architectures, a TopoPointNet (TPN) and Boosted Decision Trees (BDT). Since event-level classification alone is insufficient to restore the critical scaling, a second stage applies a particle-level density filter, explicitly stripping away the diffuse thermal background and isolating the densely packed critical clusters. The two stage pipeline successfully restores the power-law scaling of the normalized factorial moments, enabling accurate recovery of the intermittency index in ($\eta$, $\varphi$) space and establishing topological machine learning as a robust data driven tool for probing the QCD critical point and the phase structure of strongly interacting matter in heavy-ion collisions at LHC energies.
\end{abstract}

%\keywords{Suggested keywords}
\maketitle

%\tableofcontents

\section{\label{sec:section1}Introduction}

In the first microseconds of the cosmos, matter existed in a form, fundamentally different from any form of matter observed today, a extremely hot, densely packed medium in which quarks and gluons moved as free, unbound constituents~\cite{Shuryak:1978ij, Niida:2021wut}. Recreating, observing and understanding this primordial state remains one of the central pursuits of high-energy physics. Central to this quest is the Standard Model of particle physics~\cite{Woithe:2017lzd}, the most complete theoretical framework we have for describing matter and its non-gravitational interactions. At its core lies Quantum Chromodynamics (QCD)~\cite{Barber:1979yr}, the theory of the strong interaction, whose behaviuor is shaped by two defining features. The first is confinement, the principle that colour-charged quarks and gluons can never be observed in isolation~\cite{Gribov:1999ui}. The second is asymptotic freedom~\cite{Gross:1973id, Gross:1973ju, Gross:1973zrg}, whereby the strong coupling constant $\alpha_s$ grows weaker as interactions occur at increasingly smaller scales or equivalently higher momentum transfers. This interplay between confinement and asymptotic freedom governs the transition of strongly interacting matter from a hadronic phase to a deconfined state under extreme conditions. Under ordinary conditions, quarks and gluons remain permanently confined within hadrons. However, at extreme temperatures and energy densities the confining potential weakens sufficiently for quarks and gluons to decouple from individual hadrons, evolving instead as quasi-free constituents over extended volumes. This gives rise to a state known as the Quark-Gluon Plasma(QGP)~\cite{PHENIX:2003pfh}.The transition between these two regimes of strongly interacting matter is captured by the QCD phase diagram~\cite{Koch:2025cog}, mapped as a function of temperature $T$ and baryon chemical potential $\mu_B$. At low $T$ and $\mu_B$, matter exists as a confined gas of hadrons. As temperature rises, lattice QCD calculations predict a smooth crossover into the QGP at a pseudo-critical temperature $T_c \approx 155$ MeV at vanishing $\mu_B$~\cite{Cheng:2006qk}. At finite $\mu_B$, however, this crossover is conjectured to terminate at a critical endpoint, beyond which the transition becomes a first-order phase transition~\cite{Odyniec:2019kfh, Heinz:2000bk}. 
%Pinpointing the location of this critical point and establishing the order of the transition remains one of the most compelling open questions in high-energy nuclear physics.
\par

One experimentally accessible avenue for probing this phase structure is through fluctuations in charged particle multiplicity~\cite{Baym:1999up, Koch:2001zn}, which serve as a sensitive observable for the properties of the QGP and in particular, its transition to hadronic matter. Such fluctuations, when exhibiting self-similar or fractal patterns across different scales, carry direct information about the proximity of the system to a critical point. Scale-invariant fluctuations in multiparticle production can be probed within the framework of Normalized Factorial Moments (NFM)~\cite{Bialas:1988wc, DeWolf:1995nyp}, $F_q(M)$, extracted from the spatial configurations of charged particles in phase space. For a system exhibiting dynamical fluctuations driven by critical behaviour near the phase transition, $F_q(M)$ exhibits a power-law growth with increasing bin number (decreasing bin size), a phenomenon termed intermittency~\cite{Bialas:1985jb, Hwa:2016khr}.

To simulate critical fluctuations related to self-similar intermittency, the Critical Monte Carlo (CMC) model~\cite{Antoniou:2006zb, Antoniou:2005am, Wu:2022aio} has been widely employed to generate event samples exhibiting scale-invariant momentum distributions. In this model, momentum distributions of final-state particles are generated using the L\'{e}vy random walk algorithm, where the probability density between two adjacent walks follows a power-law distribution governed by the L\'{e}vy exponent $\mu = 1/6$, corresponding to a critical system belonging to the 3D Ising universality class. The CMC model thus serves as a powerful theoretical benchmark, bridging the gap between the predictions of critical phenomena and experimental observations in heavy-ion collisions.

Intermittency analyses have been carried in the two-dimensional transverse momentum ($p_x$-$p_y$) space across a wide range of center-of-mass energies. Experimentally, intermittency was first observed in central Si+Si collisions at the maximum SPS energy of 158$A$ GeV by the NA49 collaboration. However, the NA61/SHINE experiment~\cite{ReynaOrtiz:2024hul, Podlaski:2024kxg} at near-SPS energies did not detect any intermittency signals in central Ar+Sc collisions at either 150$A$ GeV or 13$A$--75$A$ GeV~\cite{NA61SHINE:2024xdd}. The STAR experiment at RHIC energies~\cite{STAR:2023jpm} observed a power-law intermittency behavior of scaled factorial moment (SFM) ratios in central Au+Au collisions, with the extracted scaling exponent displaying a non-monotonic energy dependence over the center-of-mass energy range $\sqrt{s_{NN}} = 7.7$--$200$ GeV. It is important to note that all of these experimental measurements were performed exclusively in the two-dimensional transverse momentum ($p_x$-$p_y$) space.
Subsequent investigations in $p_x$-$p_y$ space revealed that the intermittency signal observed by NA49 can be reproduced by a mixed sample consisting of 99\% background random tracks and only 1\% signal particles generated from the CMC model. Similarly, for NA61/SHINE data, the upper limit on the signal particle fraction was found to be around 1\%, and the STAR results are consistent with a mixture of approximately 1--2\% CMC signal particles embedded in a UrQMD background. These findings collectively suggest that even if critical fluctuations exist in heavy-ion collisions, the signal is extremely weak, constituting only a few percent of the entire event sample, and is easily overshadowed by the dominant non-critical background. Furthermore, when the intermittency index $\phi_2$ is calculated directly from such weak signal events in $p_x$-$p_y$ space, the dominant background particles obscure and distort the contribution of the signal particles, leading to a significant underestimation of the true intermittency index.

In parallel, an alternative and complementary approach has emerged, based on spatial fluctuations in the $(\eta, \varphi)$ phase space. This approach offers distinct practical advantages in the context of modern detector experiments, as $\eta$ and $\varphi$ are directly accessible observables reconstructed from charged particle tracks, making the analysis more naturally aligned with the detector geometry and acceptance of experiments such as ALICE at the LHC~\cite{ALICE:2014sbx}. Within the ALICE experiment, intermittency analyses have been performed by studying the NFM of charged particle multiplicity distributions in the $(\eta, \varphi)$ phase space in Pb--Pb collisions at $\sqrt{s_{NN}} = 2.76$ TeV~\cite{Sharma:2023ndr} and $\sqrt{s_{NN}} = 5.02$ TeV~\cite{Malik:2024ltm}, reporting signatures of scale-invariant fluctuations consistent with critical behavior near the QCD phase transition, with the extracted scaling exponent $\nu$ found to be in agreement with Ginzburg-Landau theory~\cite{Hwa:1992uq} predictions within experimental uncertainties. However, despite these advances, the challenge of weak signal extraction persists in the $(\eta, \varphi)$ phase space as well, and the CMC model, which has served as the standard theoretical benchmark for generating and identifying critical fluctuations in ($p_x$-$p_y$) space has not yet been extended to this phase space. To address this challenge, machine learning (ML) has emerged as a powerful data-driven approach for extracting the weak critical signal from the overwhelmingly dominant non-critical background, offering pattern recognition capabilities that can identify subtle structures in complex, high-dimensional datasets that traditional factorial moment analysis alone cannot resolve. In particular, topological machine learning, which combines Topological Data Analysis (TDA) with deep learning architectures has recently demonstrated remarkable success in classifying weak intermittency signal events from background noise in $p_x$-$p_y$ space, by extracting distinct topological features of the particle distributions through persistent homology, using the CMC model as the source of injected signal events.

In the present work, we address this gap by implementing the CMC-based approach in the $(\eta, \varphi)$ phase space for the first time. By incorporating the CMC model as the source of critical signal events and the EPOS model as the representative background, we employ a topological machine learning framework to classify signal events from background, extract the weak intermittency signal, and accurately determine the intermittency index in this phase space. This approach allows us to directly assess the impact of the CMC signal in the $(\eta, \varphi)$ phase space and provides a more complete and robust picture of critical fluctuations in heavy-ion collisions at LHC energies. To the best of our knowledge, this is the first application of the CMC model combined with topological machine learning in the $(\eta, \varphi)$ phase space, opening a new avenue for the study of intermittency and critical phenomena in ultra-relativistic heavy-ion collisions.

\par
This paper is organized as follows. 
In Sec.~\ref{sec:section2}, we provide a brief overview of the intermittency framework and the Critical Monte Carlo (CMC) model. Sec.~\ref{sec:section3} describes the event samples used in this analysis and the signal embedding procedure, including a discussion of the effect of signal dilution on the embedded critical fluctuations. In Sec.~\ref{sec:section4}, we introduce the persistent homology framework employed in this work, covering the Delaunay filtration construction, the extraction of Betti curves, and the effect of azimuthal randomisation on the Betti curve distributions. Sec.~\ref{sec:section5} presents the machine learning classification of topological features using two complementary approaches: the TopoPointNet (TPN) and Boosted Decision Trees (BDT). The results of the analysis are presented and discussed in Sec.~\ref{sec:section6}.

\section{\label{sec:section2}Intermittency and Critical Monte Carlo (CMC) model}
To quantify local multiplicity fluctuations and identify scale-invariant properties, the method of normalized factorial moments (NFMs) introduced by Bialas and Peschanski is employed~\cite{Bialas:1985jb,Bialas:1988wc}. A primary advantage of NFMs is their ability to filter out purely statistical Poissonian noise associated with finite particle numbers, isolating genuine dynamical correlations. For a two-dimensional phase space spanned by pseudo-rapidity $\eta$ and azimuthal angle $\varphi$, the acceptance region is partitioned into a uniform $M \times M$ grid yielding $M^2$ total cells. The $q$-th order normalized factorial moment \fq{q} is defined as 
\begin{equation}
  F_{\rm{q}}(M)= \frac{\frac{1}{N} \displaystyle\sum_{e=1}^{N}\frac{1}{M^2}\sum_{i=1}^{M}f_{q}^{e}(n_{\rm{ie}})}{\left (\frac{1}{N} \displaystyle \sum_{e=1}^{N}\frac{1}{M^2}\sum_{i=1}^{M}f_{1}^{e}(n_{\rm{ie}}) \right )^q},
\label{eq1}
\end{equation}
 \par
 where,
 \begin{equation}
 f_{q}^{e}(n_{\rm{ie}}) = \displaystyle\prod_{j=o}^{q-1}(n_{\rm{ie}} - j)
\label{equation2}
\end{equation}
where \two{n}{ie} represents the particle multiplicity in the $i^{\rm{th}}$ cell of the $e^{\rm{th}}$ event.
In the absence of dynamical fluctuations, the multiplicity distribution follows Poisson statistics and \fq{q} remains independent of the partition scale $M$. Conversely, if the system features self-similar, scale-invariant density fluctuations the SFMs exhibit a power-law growth with respect to the number of cells at large $M$. This behaviour, termed intermittency~\cite{Bialas:1985jb,Bialas:1988wc} is characterized as: 
\begin{equation} 
F_{\rm{q}}(M) \propto M^{\phi_{q}},
\label{eq2}
\end{equation}
where \two{\phi}{q} is the $q-$th intermittency index. A strictly positive \two{\phi}{q} $> 0$ signals the presence of scale-invariant correlations and fractal particle clustering, whereas \two{\phi}{q} $\approx 0$ indicates a purely thermal, uncorrelated background. 
% The intermittency index connects directly to the multifractal structure of the emission source, allowing the extraction of the multifractal dimension Dq. 
In general, the intermittency is evaluated by examining the dependence of the NFMs on the number of phase-space cells using Eq. \ref{eq2} in a logarithmic representation. The power-law behaviour of NFMs takes a linear form with the slope \two{\phi}{q}. For a system near a phase transition, scale invariance is an expected consequence of diverging correlation lengths, producing a distinctive fractal signature in the final-state angular distribution. If the clustering originates from a second-order phase transition in the 3D Ising universality class, theoretical calculations predict a second-order intermittency index of \two{\phi}{2} = 2/3 in a two-dimensional momentum space. It should be noted that the value is predicted for (\two{p}{x}, \two{p}{y}) space and not ($\eta, \varphi$) space~\cite{Antoniou:1998np}.

To model the explicit scale-invariant fluctuations, the Critical Monte Carlo (CMC)~\cite{Antoniou:2000ms} algorithm is utilized. The CMC framework generates particle configurations with a well-defined multifractal geometry through a Lévy random walk. Unlike a standard Brownian walk, which yields a uniform macroscopic density and a trivial monofractal dimension, the Lévy walk allows for heavy-tailed step sizes. This mechanism naturally produces nested, self-similar clusters across multiple length scales, simulating the diverging correlation length at a phase transition.
The probability density function, $P(r)$ for the step size, $r$ between consecutively generated particles is governed by a power-law distribution, 
\begin{equation} P(r) = \mu r_{\text{min}}^\mu \left[ 1 - \left( \frac{r_{\text{min}}}{r_{\text{max}}} \right)^\mu \right]^{-1} r^{-1-\mu},
\end{equation}
where \two{r}{min} and \two{r}{max} define the lower and upper bounds of the step size. In contrast to original CMC implementation~\cite{Antoniou:2000ms}, where the Lévy walk is defined in (\two{p}{x}, \two{p}{y}) space, this work performs the walk directly in the $(\eta, \varphi)$ acceptance window. The upper bound is set to $\mtwo{r}{max} = 2\pi$, corresponding to the full azimuthal extent. The lower cutoff, \two{r}{min} $= \mtwo{r}{max} \times 10^{-7}$, serves solely as a numerical regularisation of the power-law divergence as $r \to 0$ and carries no physical significance at the relevant track-separation scales. These boundary values establish the explicit dynamic range over which the fractal geometry is mathematically exact. The scaling properties of the resulting point cloud are uniquely dictated by the Lévy exponent $\mu$. For a critical system belonging to the 3D Ising universality class, theoretical calculations constrain this exponent to $\mu =$ 1/6. By iteratively sampling radial steps from this heavy-tailed distribution and drawing azimuthal directions uniformly, one constructs an isolated fractal cluster. This localized signal can subsequently be injected into a non-critical background to isolate and study the purely dynamical fluctuations against a complex thermal medium.
All previous applications of the CMC algorithm have predominantly formulated the Lévy random walk in the cartesian momentum space (\two{p}{x}, \two{p}{y}). In that flat Euclidean geometry, the generation of critical clusters uniquely recovers the theoretically predicted \two{\phi}{2} = 2/3. The present work follows the methodology introduced by Hwa and Yang~\cite{Hwa:2011bu}, which formulates the intermittency analysis directly within the ($\eta, \varphi$) plane. The same approach is also followed in various phenomenological and experimental works in~\cite{Sharma:2018vtf,Gupta:2019zox,Haider:2026rtw,Sarma:2019teo,Singh:2024gai,Sharma:2023ndr,Malik:2024ltm}. Similar investigations in the ($\eta, \varphi$) space have been performed previously using Toy MC at \energy{2.76} that samples ALICE multiplicity distributions~\cite{Sharma:2023oxo}, however, the study models, purely thermal backgrounds without implementing any CMC algorithm to inject explicit scale-invariant critical dynamics.

\section{\label{sec:section3}Event sample and signal embedding}

EPOS is a hybrid event generator for high-energy $pp$ and $AA$ collisions that combines perturbative QCD, Gribov–Regge theory and relativistic viscous hydrodynamics within a unified framework~\cite{Werner:2023mod, Werner:2024fwk, Werner:2023zvo, Bass:1998ca}. Initial-state partonic interactions are modelled via multiple Pomeron exchanges with colour-saturation effects from the Colour Glass Condensate formalism~\cite{McLerran:1993ka}, followed by a core–corona separation in which the dense thermalised core evolves hydrodynamically to form the QGP while the dilute corona decouples early. Hadronisation proceeds via a combination of coalescence and fragmentation, reproducing bulk observables including strangeness enhancement, collective flow and identified-particle \two{p}{T} spectra across collision energies up to the LHC scale. Intermittency has been studied with earlier EPOS versions and other models at LHC and RHIC energies~\cite{Gupta:2019zox,Wu:2021jou,Haider:2026rtw,Wu:2022aio} and no statistically significant power-law scaling has been observed in the simulated samples. This make EPOS a well-validated baseline for the present study, in which its events serve as the inert background for this analysis into which the CMC signal is injected.

$6 \times 10^5$ minimum bias events are generated for Pb--Pb collisions at \energy{5.02}. Charged-particle multiplicity in the V0 detector acceptance, V0M, is used to  classify events into centrality classes. The V0 detector comprises two arrays, V0A ($2.8<\eta<5.1$) and V0C ($-3.7<\eta<-1.7$), and the V0M amplitude is estimated by  counting charged particles within this acceptance~\cite{ALICE:2013axi}. Events are ordered by their V0M multiplicity, the resulting distribution is divided into percentiles that serve as a proxy for centrality~\cite{ALICE:2018tvk}. The analysis focuses exclusively on the most central collisions (0--5\%) for maximum track multiplicity. From these generated collisions, we construct two parallel datasets each with $3 \times 10^5$ events to train and validate the machine learning pipeline. The first dataset serves as the background reference and consists entirely of pure EPOS events. Using this first dataset, signal event samples are constructed from the exact same underlying EPOS host events after being subjected to the CMC injection. After applying the centrality constraint, the dataset yields approximately $15\times 10^3$ events per dataset. The analysis is performed on final-state charged hadrons ($\pi^{\pm}$, $K^{\pm}$, $p \overline{p}$ )  within the pseudo-rapidity window $|\eta| <$ 0.8 and the transverse momentum range 0.2 $< \mtwo{p}{T} <$ 3.0 GeV/\textit{c}. This \two{p}{T} interval predominantly isolates the bulk soft-particle production.

\begin{figure}[h!]
\includegraphics[width=0.49\textwidth]{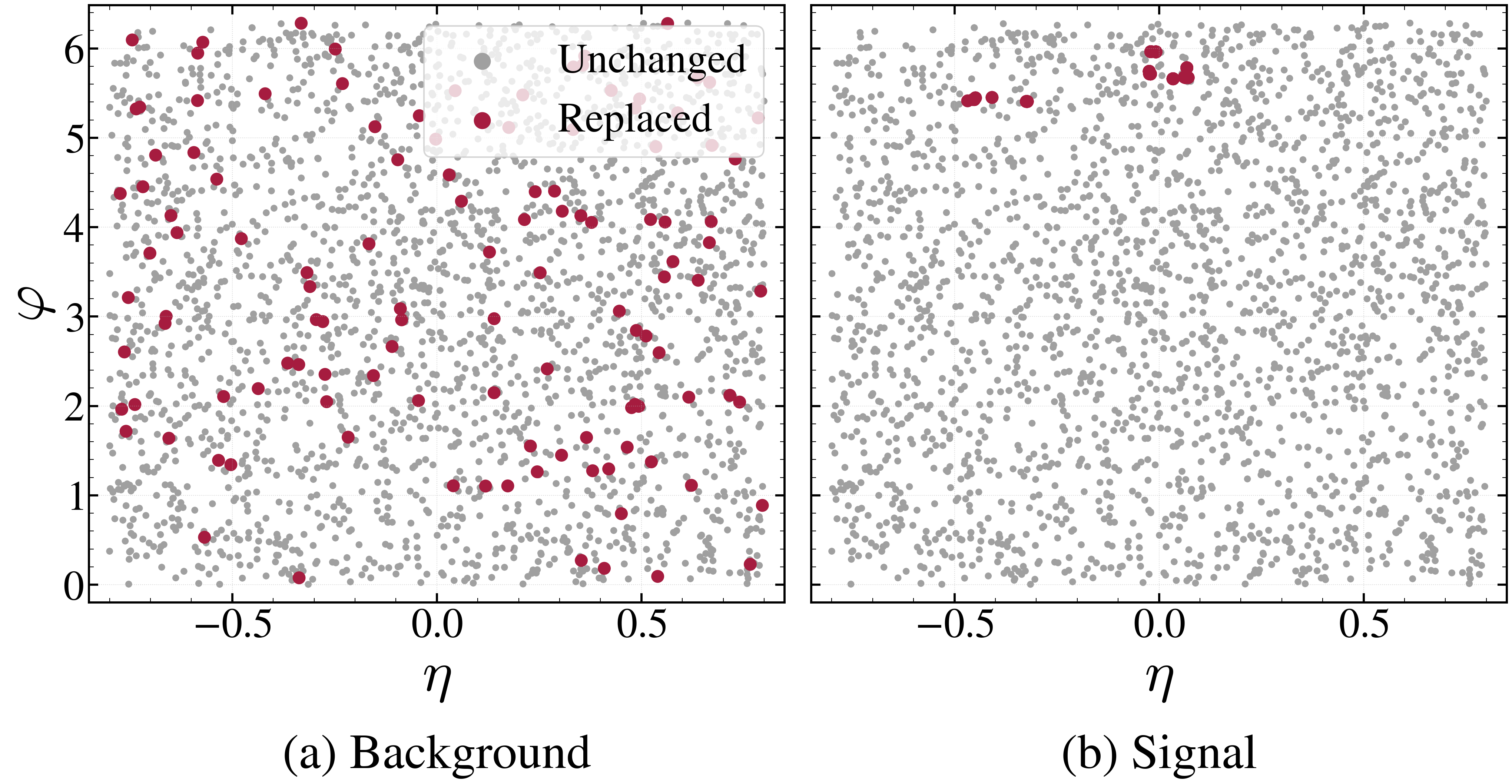}
\caption{\label{fig1} Two-dimensional spatial track distribution in the $(\eta, \varphi)$ phase space for 0--5\% central Pb--Pb collision at \energy{5.02}. Unmodified EPOS tracks (grey) are shown alongside 5\% embedded CMC signal tracks (red) within $|\eta| < 0.8$ and $0.2 \le \mtwo{p}{T} \le 3.0\text{ GeV}/c$, illustrating localized critical clusters (right).}.
\end{figure}

\begin{figure*}[htb]
    \centering
    \includegraphics[width=0.48\textwidth]{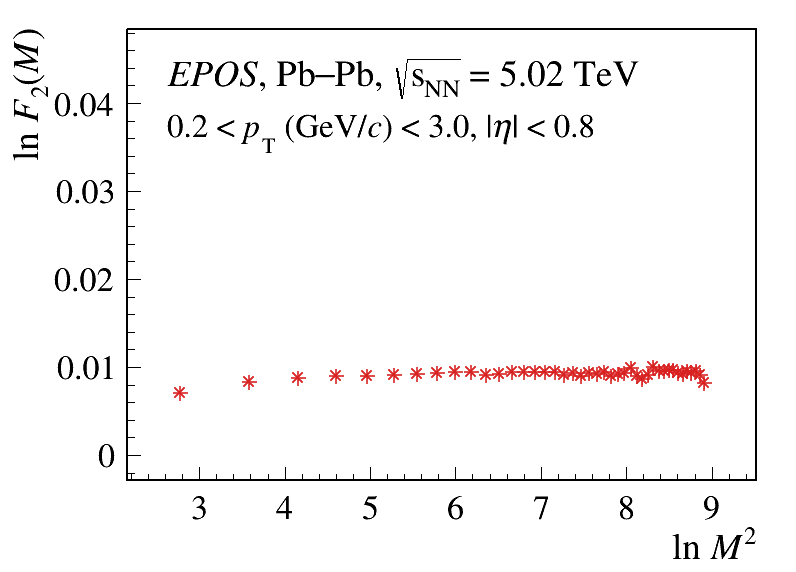}
    \includegraphics[width=0.48\textwidth]{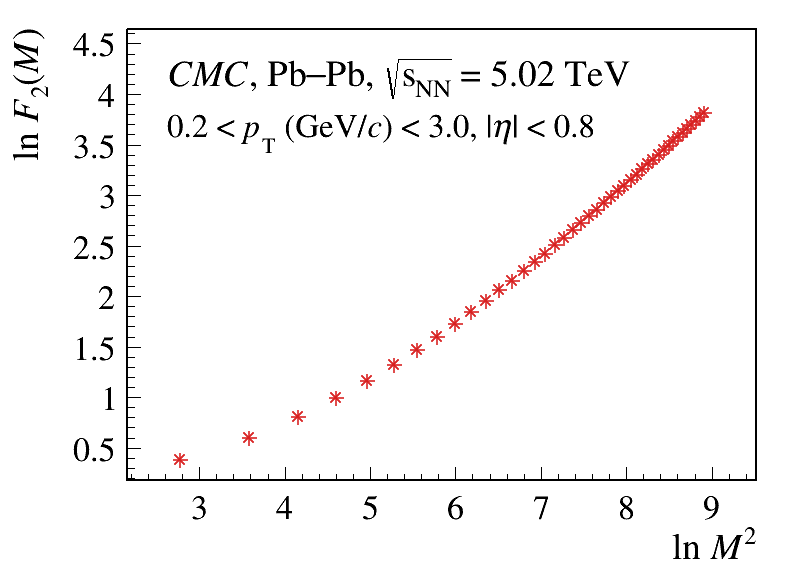}
    \caption{\label{fig2} Log-log dependence of the second-order NFM, \fq{2} on $M^2$ for EPOS (left) and CMC ($\lambda = 1.0$, right) in 0--5\% central Pb--Pb collisions at \energy{5.02}.}
\end{figure*}

To embed the signal, particles are replaced in the generated EPOS events. Within each event, a fixed fraction, $\lambda$ of the accepted final-state tracks is randomly selected for modification. For these selected tracks, a localized CMC Lévy walk is generated directly in the angular $(\eta, \varphi)$ plane. The original angular coordinates of the selected EPOS tracks are subsequently replaced by the generated CMC coordinates, while their original \two{p}{T} is preserved exactly. The Cartesian momentum components are recomputed as:
\begin{equation} 
p_{\rm{x}} = p_{\rm{T}} \cos\varphi, \quad p_{\rm{y}} = p_{\rm{T}} \sin\varphi, \quad p_{\rm{z}} = p_{\rm{T}} \sinh\eta, 
\end{equation}
ensuring exact consistency between the kinematic and angular variables. The scale-invariant fluctuations are thus hidden exclusively within the many-particle angular correlations. The resulting angular distribution of tracks for a single event is shown in Figure \ref{fig1}. The 2D scatter shows the unmodified EPOS background tracks (grey) alongside the replaced CMC signal tracks (red). In the signal event, the CMC-replaced tracks form tight spatial clusters, which initially were distributed uniformly across the acceptance.

\begin{figure}[h!]
    \includegraphics[width=0.49\textwidth]{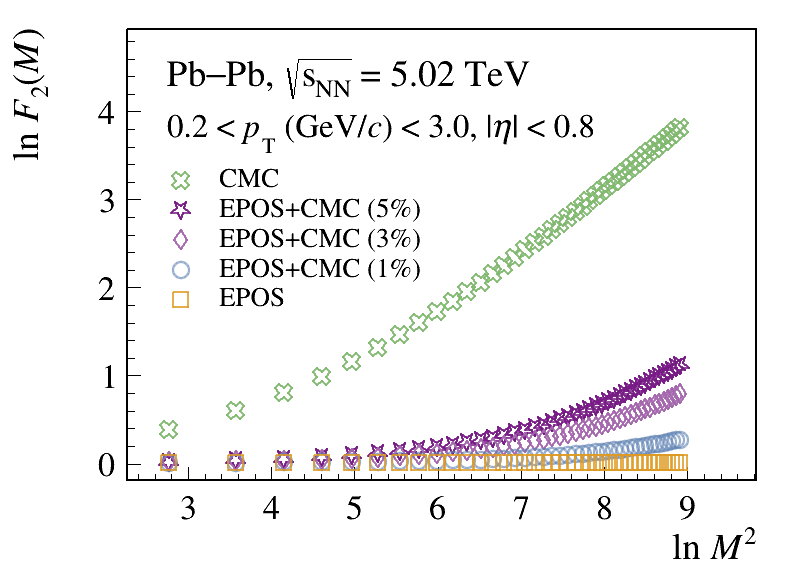}
    \caption{\label{fig2a} Scaling dilution of second-order NFM, \fq{2} as a function of $M^2$ in $(\eta, \varphi)$ space. Markers are shown for pure CMC ($\lambda = 1.0$), diluted mixtures ($\lambda = 0.05, 0.03, 0.01$), and pure EPOS background in 0--5\% central Pb--Pb collisions at \energy{5.02}.}
\end{figure}

The above procedure yields different parallel event samples constructed from the same EPOS events. The first are the unmodified background events ($\lambda = 0$), and the second are the $X$\% signal mixture ($\lambda = 0.0X$) into background events. The third is a pure CMC sample ($\lambda = 1.0$) generated within the identical kinematic acceptance.

\subsubsection{\label{sec:section3.1}Dilution in the embedded signal}
The section details the sensitivity of intermittency to the embedded signal. NFMs are calculated for the event samples and shown in Figure \ref{fig2} as a log-log dependence of \fq{2} on the number of phase-space bins $M^2$. For the pure EPOS background (left), \fq{2} exhibit no significant dependence on the increasing phase space bins. The scaling behaviour of the pure CMC reference sample (right) demonstrates strict power-law growth. Figure \ref{fig2} establishes the baseline scaling behaviour of the background and pure signal. Statistical uncertainties on \fq{2} are calculated using the sub-sampling method.

Figure \ref{fig2a} shows the sensitivity of intermittency to the concentration of the embedded critical signal. It compiles the scaling behaviours of different signal mixtures including 1\% ($\lambda = 0.01$), 3\% ($\lambda = 0.03$) and 5\% ($\lambda = 0.05$) signal replacement ratios compared with pure background and pure signal. However, the signal mixtures shows a diminished power-law compared to pure CMC.   At low $M^2$, the signal mixtures remain indistinguishable from the background. A departure from the background scaling emerges only at higher $M^2$. Because the vast majority of accepted tracks originate from the uniform background, the macroscopic event characteristics heavily dilute and suppress the localized critical clusters. It is interesting to observe that the embedding of 5\% signal gives \fq{q} trend similar to that observed in ToyMC~\cite{Sharma:2023oxo}. 
%\textcolor{red}{check in cmc paper, whether their sensitivity is low or high compared to us, and then write a sentence or two mentioning that.}
 
\begin{figure}[h!]
    \includegraphics[width=0.49\textwidth]{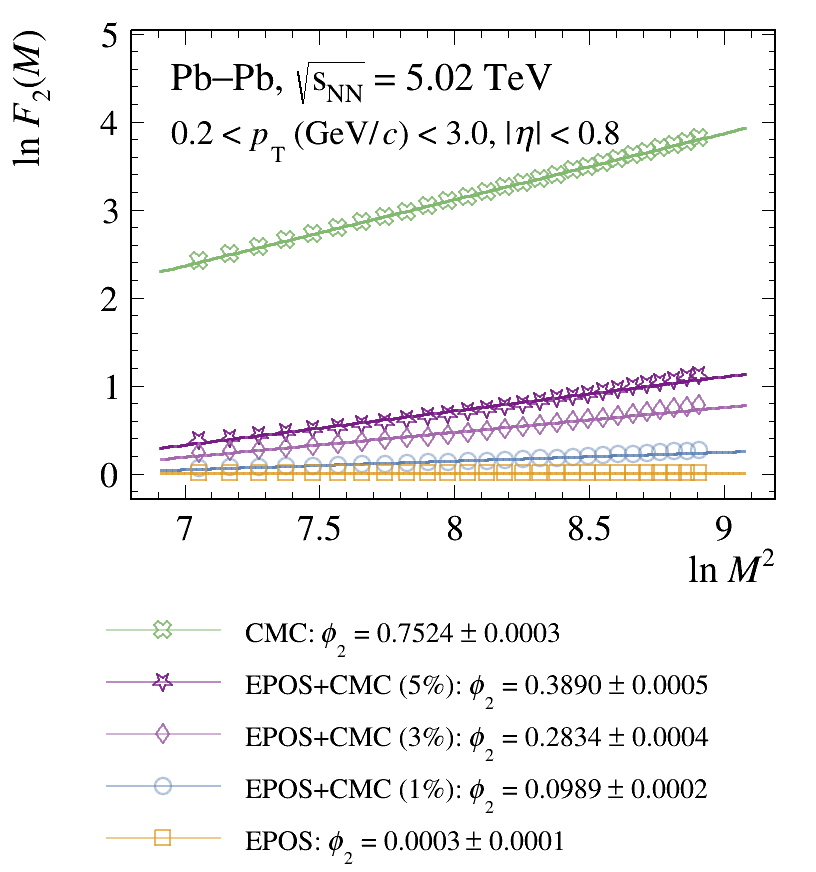}
    \caption{\label{fig3}  Linear fits shown in second-order NFM, \fq{2} vs $\ln M^2$ plot to extract intermittency index, \two{\phi}{2} across signal replacement fractions $\lambda$ in 0--5\% central Pb--Pb collisions at \energy{5.02}. Fits are performed in higher $M^2$ region ($\in [6.9, 9.1]$). Pure CMC gives $\phi_2 \approx 0.73$, whereas all diluted samples ($\lambda \le 0.05$) collapse toward the background value ($\phi_2 \approx 0$), demonstrating signal suppression in raw phase-space moments.}
\end{figure}

The resulting intermittency index, \two{\phi}{2} is calculated for all curves shown in Figure \ref{fig2a}, by fitting the linear regions and are shown in Figure \ref{fig3}. For pure CMC, a shifted value of \two{\phi}{2} $\approx 0.73$ from the theoretically predicted value $\mtwo{\phi}{2} \approx 2/3 (= 0.667)$ is observed. This shift in value may be the direct geometric consequence of the $(\eta, \varphi)$ coordinate transformation and the finite-size acceptance boundaries. Nevertheless, these values of \two{\phi}{q} for pure CMC will serve as a reference for ours and any future analysis performed in this geometry. It can be seen that EPOS+CMC values 1\%, 3\% and 5\% signal mixtures are heavily suppressed,  lying substantially closer to EPOS than to CMC. This suppression grows as the signal fraction decreases, with $\mtwo{\phi}{2}$ shifting progressively toward the background value.

The dilution of power law in EPOS+CMC (Figure \ref{fig2a}) and the value of \two{\phi}{2} in EPOS+CMC being away from CMC (Figure \ref{fig3}) shows that the application of NFMs is insufficient to extract a weak critical signal from a dominant background. To recover the underlying scale-invariant dynamics, the critical clusters must be geometrically identified and separated from the thermal bulk prior to the factorial moment calculation. This dilution explicitly motivates the implementation of a topological filtering pipeline to achieve discrimination at the event and more importantly particle levels. The capacity of the subsequent topological pipeline is evaluated by its ability to recover the specific configuration-dependent baseline from a dilute background mixture.

%This pure sample is not utilized in the machine learning training; rather, it is constructed solely to evaluate the shifted numerical baseline for the intermittency index in the finite (η,φ) geometry, establishing the absolute mathematical reference for signal recovery. For all statistical evaluations of the scaled factorial moments across these datasets, uncertainties are estimated via a sub-sampling method, dividing the full event classes into ensembles of 100 events to extract the standard error of the mean

\section{\label{sec:section4}Persistent homology framework}
The dilution of the embedded signal highlights a fundamental limitation of standard NFMs~\cite{Hwa:2011bu,NA49:2012ebu}. Traditional intermittency analysis relies on partitioning the phase space into rigid grids and evaluating local bin multiplicities. This histogram-based approach is sensitive to boundary placements and invariably averages out localized structural information across the full experimental acceptance. To bypass these limitations, Topological Data Analysis (TDA) is used in this work. It evaluates the intrinsic shape and connectivity of the data rather than relying on discrete summary statistics~\cite{Leykam:2022ejk}.
The final-state particle kinematics of a heavy-ion collision naturally form a discrete point cloud in the $(\eta, \varphi)$ space. TDA provides a coordinate-free framework to characterize the multi-particle correlational structure of such spatial distributions~\cite{Hamilton:2022blu}. The Lévy walk in CMC generates tightly localized clusters, which manifest as distinct topological objects. These critical clusters exhibit a characteristic geometry, connected components and localized loops that appear, merge and close at specific spatial scales. EPOS background lacks this intricate connectivity, filling the phase space with a uniform distribution. TDA isolates the geometry of the critical clusters from the overwhelming background by tracking the topological evolution of the point cloud. This is done first by defining a continuous geometric filtration of the data and subsequently extracting scale-dependent topological invariants known as Betti curves~\cite{Hamilton:2022blu}.

\begin{figure*}[htb]
    \includegraphics[width=0.99\textwidth]{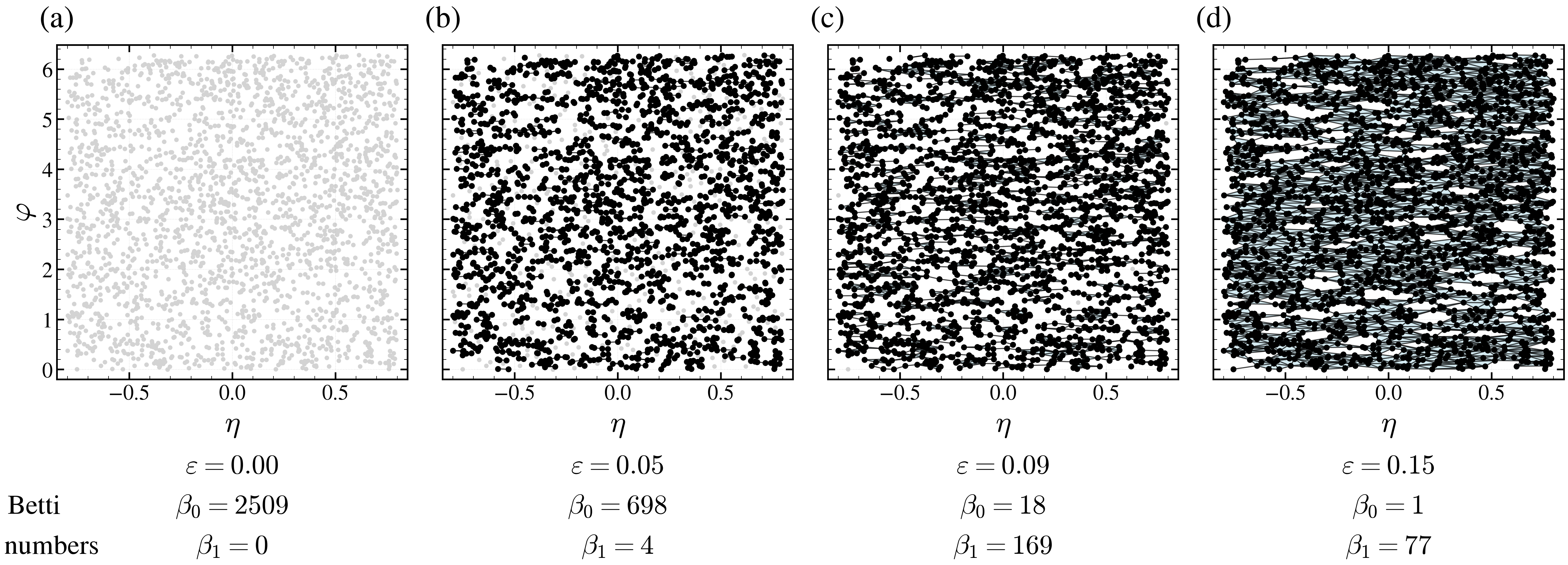}
    \caption{\label{fig4} Evolution of the 2D periodic Delaunay filtration in $(\eta, \varphi)$ space of charged particles in a Pb--Pb collision event at \energy{5.02} generated by EPOS across scale thresholds $\varepsilon$. As $\varepsilon$ increases, active vertices ($d_{\mathrm{NN}} \le \varepsilon$), edges and triangles progressively form simplicial subcomplexes, tracking the birth and collapse of connected components ($\beta_0$) and 1D loops ($\beta_1$).}
\end{figure*}

\subsubsection{\label{sec:section4.1}Delaunay filtration}
In order to analyse the multi-particle correlations within an event, topology relies on the construction of a simplicial complex. A simplicial complex generalizes a standard network graph by incorporating higher-dimensional geometric structures, for a two-dimensional point cloud, this comprises vertices (0-simplices), edges (1-simplices), and triangular faces (2-simplices)~\cite{Capellino:2025kce}. Discrete final-state kinematics are mapped into this framework using the Delaunay triangulation. For a given set of points, the Delaunay partitions the space into triangles such that no point lies within the circumcircle of any triangle~\cite{Hamilton:2022blu,Wang:2024bzy}. This circumcircle optimality maximizes the minimum interior angle of all generated triangles, providing a geometrically well-conditioned representation of the local connectivity. It establishes a nonparametric definition of spatial proximity, particles are considered adjacent if and only if they are connected by a 1-simplex in the Delaunay complex. The application of this triangulation to collider data requires a specific geometric correction. The azimuthal angle $\varphi$ imposes periodic boundary conditions on the interval [0, $2\pi$]. A naive triangulation evaluated in a flat Cartesian plane fails to connect physically adjacent tracks separated by the $\varphi=0$ boundary, artificially disrupting the topology of macroscopic clusters. To rigorously enforce the cylindrical metric of the $(\eta, \varphi)$ acceptance, we employ a phantom-point duplication scheme~\cite{Hamilton:2022blu}. For every primary track $i$ recorded at $(\mtwo{\eta}{i}, \mtwo{\varphi}{i})$, two identical phantom copies are projected at $(\mtwo{\eta}{i}, \mtwo{\varphi}{i} - 2\pi)$ and $(\mtwo{\eta}{i}, \mtwo{\varphi}{i}+2\pi)$ excluding phantom copies of the same particle $i$. The Delaunay triangulation is subsequently computed across the augmented dataset of 3N particles. Following the triangulation, any simplices composed exclusively of phantom points or those connecting two copies of the same primary track are discarded as degenerate. The retained structure forms an exact, boundary-free simplicial complex on the $(\mtwo{\eta}{i}, \mtwo{\varphi}{i})$ cylinder.

While the triangulation encodes the global connectivity of the event, extracting scale-dependent topology requires evaluating how this connectivity evolves across spatial resolutions. To achieve this, a continuous geometric filtration of the simplicial complex is constructed based on the local spatial density of the tracks. The fundamental metric for this filtration is the nearest-neighbour distance, \two{d}{NN}$(i)$ for each particle $i$, defined~\cite{Wang:2024bzy} in the $(\eta, \varphi)$ plane as: $d_{\rm{NN}}(i) = \min_{j \neq i} \sqrt{(\eta_i - \eta_j)^2 + (\varphi_i - \varphi_j)^2},$ where the distance is evaluated across the augmented phantom-point dataset to correctly preserve the azimuthal periodicity. This nearest-neighbour distance maps inversely to local particle density, tracks embedded within tightly packed critical clusters yield small \two{d}{NN} values, whereas isolated tracks in the thermal bulk yield large values. Using this metric, a sub-level-set filtration is defined to systematically probe the multi-particle correlations. A filtration scale parameter, $\varepsilon$ is swept from $0$ to a maximum radius of $\mtwo{\varepsilon}{max} = 0.5$ rad in $300$ discrete steps. At each threshold $\varepsilon$, an activation rule governs the formation of the subcomplex. A vertex (particle) $i$ is declared active if its nearest-neighbour distance satisfies $\mtwo{d}{NN}(i) \leq \varepsilon$. Higher-dimensional structures in the Delaunay triangulation, specifically edges (1-simplices) and triangular faces (2-simplices) become active if and only if all of their constituent vertices are active. This specific sub-level-set construction guarantees a strictly nested sequence of subcomplexes. As $\varepsilon$ increases, the filtration progressively incorporates particles in order of decreasing local density. The visual evolution of this sequence for a single event is illustrated in Figure \ref{fig4}. At small $\varepsilon$, only the highly correlated particles comprising the dense Lévy clusters activate. As the scale parameter grows, the uniform thermal background tracks gradually emerge and establish connections between the isolated components, eventually recovering the complete Delaunay complex over the entire kinematic acceptance~\cite{Wang:2024bzy}.

\begin{figure}[h!]
    \includegraphics[width=0.49\textwidth]{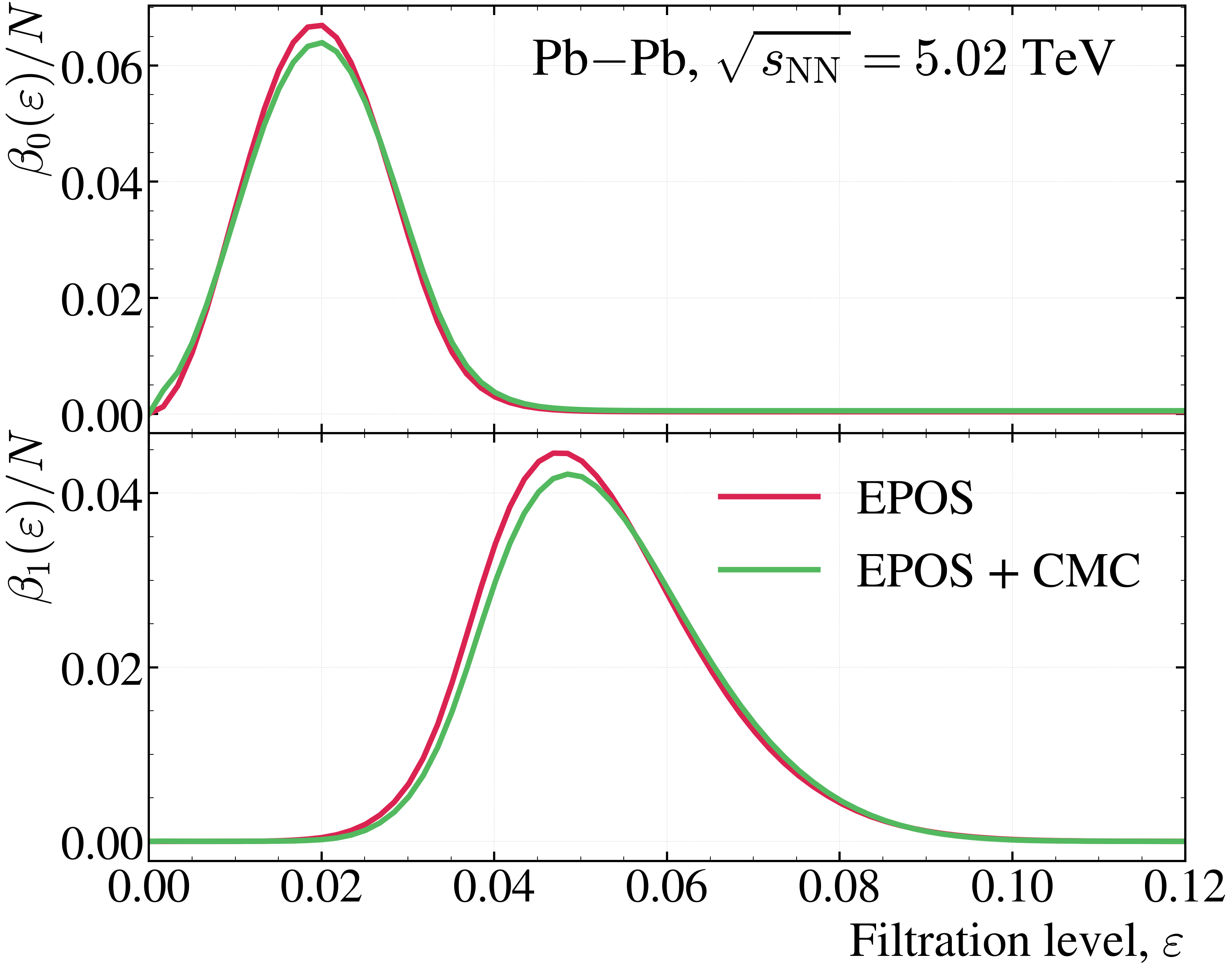}
    \caption{\label{fig5} Multiplicity-normalized raw Betti curves $\beta_0(\varepsilon)/N$ and $\beta_1(\varepsilon)/N$ vs filtration scale $\varepsilon$ for EPOS and 5\% CMC signal mixture ($\lambda = 0.05$) in central Pb--Pb collisions at \energy{5.02}.}
\end{figure}

\begin{figure*}[htb]
    \centering
    \includegraphics[width=0.49\textwidth]{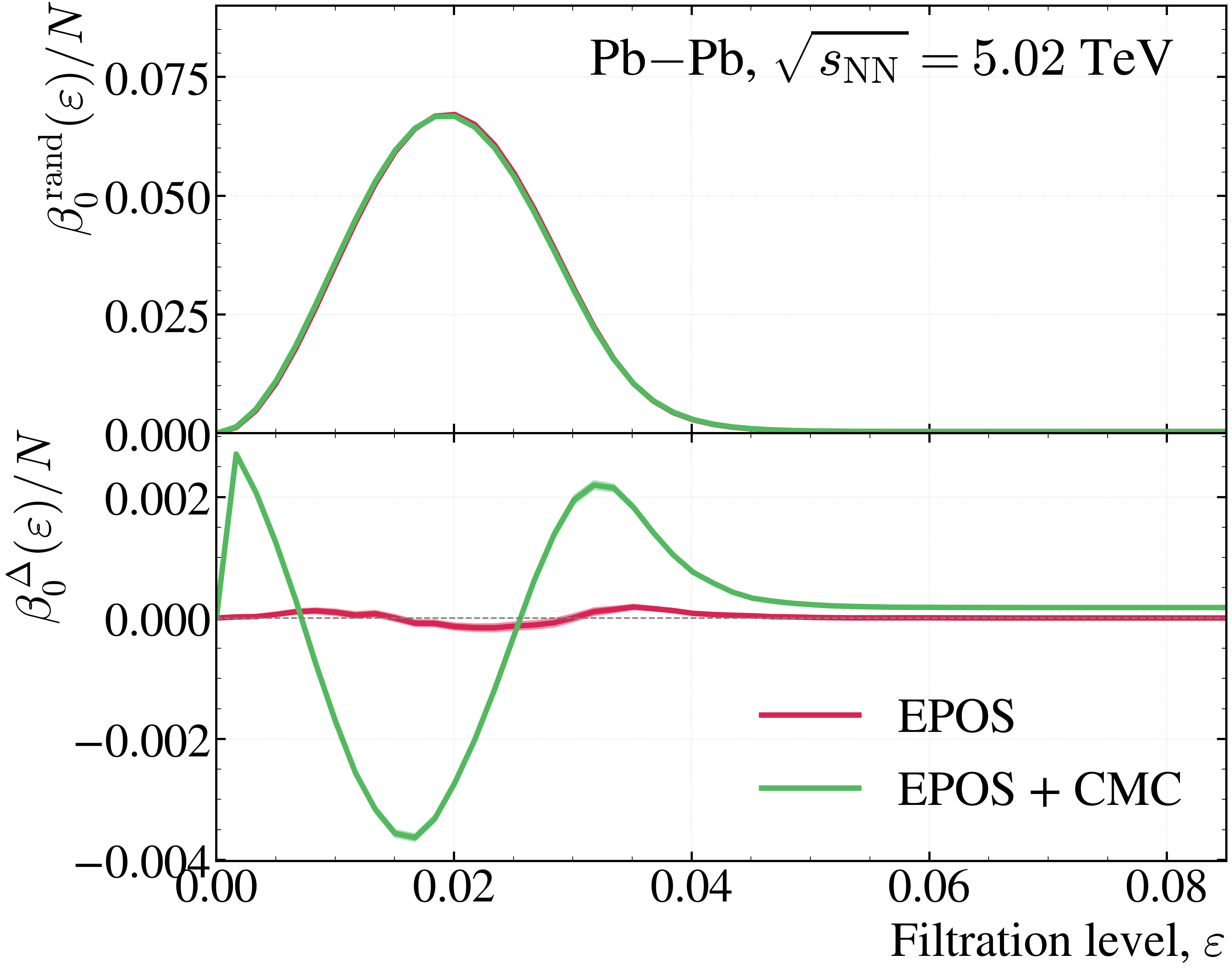}
    \includegraphics[width=0.49\textwidth]{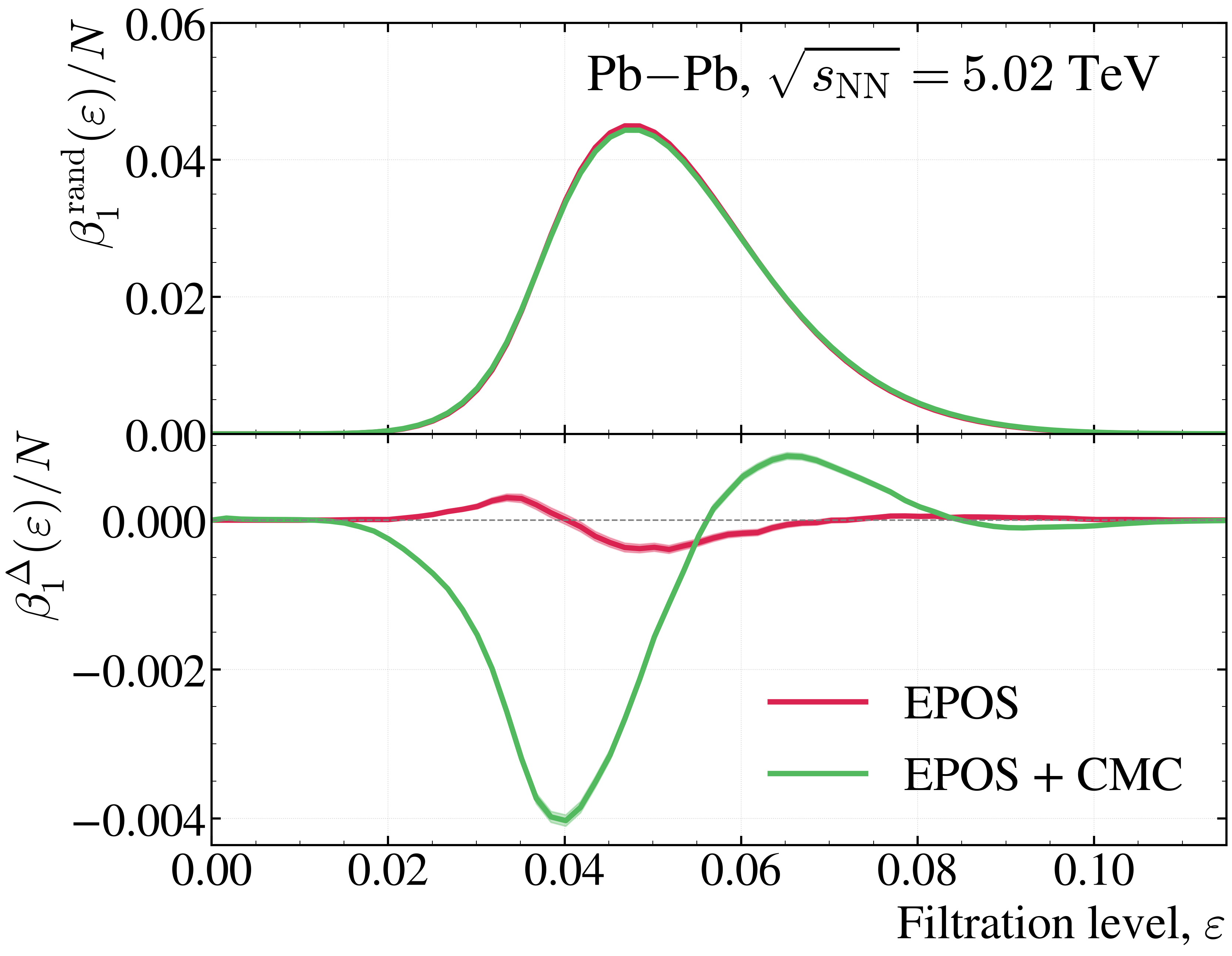}
    \caption{\label{fig6} Azimuthally randomized differential Betti curves $\beta_0^\Delta(\varepsilon)$ (left) and $\beta_1^\Delta(\varepsilon)$ (right) vs filtration scale $\varepsilon$ for pure EPOS and 5\% CMC mixture ($\lambda = 0.05$) in central Pb--Pb collisions at \energy{5.02}. The negative dip in $\beta_0^\Delta(\varepsilon)$ at small $\varepsilon \approx 0.02$--$0.05\text{ rad}$ and $\beta_1^\Delta(\varepsilon)$ at $\varepsilon \approx 0.03$--$0.05\text{ rad}$ isolate the spatial clustering signature of CMC.}
\end{figure*}

\subsubsection{\label{sec:section4.2}Extraction of Betti curves}
The sequence of subcomplexes generated by the sub-level-set filtration is topologically characterized by its homology groups. The ranks of these homology groups are the Betti numbers, which quantify the distinct topological features present at a given spatial resolution. For the two-dimensional $(\eta, \varphi)$ phase space, the relevant invariants are the zeroth Betti number $\beta_0$, which counts the number of independent connected components, and the first Betti number $\beta_1$, which measures the number of non-homologous 1D loops. As the filtration parameter $\varepsilon$ increases, the continuous evolution of these invariants defines the Betti curves $\beta_0 (\varepsilon)$ and $\beta_1 (\varepsilon)$~\cite{Hamilton:2022blu,Wang:2024bzy}. This evolution is also depicted in Figure \ref{fig4}.

Extracting Betti numbers for a general simplicial complex typically requires a computationally intensive algebraic reduction of boundary matrices. However, the Delaunay filtration constructed on the $(\eta, \varphi)$ plane is strictly 2D, comprising only vertices $(V)$, edges $(E)$ and triangular faces $(F)$ with no higher-dimensional simplices. Thus, the topological invariants are exactly constrained by the Euler–Poincaré formula~\cite{Duy:2016}. This relation equates the alternating sum of the simplex counts to the alternating sum of the Betti numbers. In 2D, this yields the Euler characteristic: $\chi = V - E + F = \beta_0 - \beta_1$~\cite{Duy:2016}. Rearranging this identity provides an exact algebraic extraction of the first Betti number, \begin{equation}
\beta_1(\varepsilon) = \beta_0(\varepsilon) - V(\varepsilon) + E(\varepsilon) - F(\varepsilon).
\end{equation} 
Here, $V(\varepsilon), E(\varepsilon)$ and $F(\varepsilon)$ represent the number of active 0-, 1-, and 2-simplices at a given filtration scale. This geometric constraint bypasses the need for matrix reduction entirely. It allows for an exact and highly efficient computation of the loop structures $\beta_1(\varepsilon)$ across the entire filtration sweep by tracking the active simplex counts alongside the connected components $\beta_0(\varepsilon)$.

The physical geometry of the event is directly encoded in the morphological evolution of the Betti curves. Figure \ref{fig5} shows dimension-0 and -1 normalized betti curves for Pb--Pb collisions at \energy{5.02} comparing background (EPOS) and signal (EPOS+CMC). At $\varepsilon = 0$, no particles satisfy the activation threshold, yielding an empty subcomplex where $\beta_0 = 0$. As $\varepsilon$ expands, particles activate and immediately begin forming connections. For events containing an embedded critical signal, the underlying Lévy walk produces tightly localized, dense spatial clusters in the $(\eta, \varphi)$ plane. These compact structures activate at very small spatial scales. Consequently, $\beta_0 (\varepsilon)$ drops steeply at small $\varepsilon$ as the densely packed vertices rapidly merge into larger connected components. Conversely, the background distributes particles more uniformly across the acceptance, resulting in a delayed and gradual decline in the number of independent components. The evolution of the first Betti number, $\beta_1 (\varepsilon)$ follow from the tight spatial proximity of the critical clusters which facilitates the early formation of triangulated loops. In the signal mixture, $\beta_1 (\varepsilon)$ rises rapidly at small $\varepsilon$ as these localized topological holes open and then falls quickly as further increases in the filtration scale fill the loops with 2-simplices. The uniform background requires substantially larger filtration scales to establish the connections necessary to form closed one-dimensional boundaries, delaying both the initial rise and the subsequent decay of $\beta_1 (\varepsilon)$.

\subsubsection{Azimuthal randomisation in betti curves}

$\beta_0 (\varepsilon)$ and $\beta_1 (\varepsilon)$ curves scale with the total number of active particles and hence, with the event track multiplicity $N$, which fluctuates across collisions even within a narrow centrality class. A machine learning classifier trained on these raw Betti curves could exploit this amplitude scaling to distinguish classes by multiplicity rather than by cluster geometry. To remove this bias, both curves are divided by $N$~\cite{Wang:2024bzy}. While this removes proportionality with $N$, the curves may still retain a systematic shape contribution from collective azimuthal flow present in the background. To explicitly subtract this, an azimuthal randomisation technique is applied~\cite{Capellino:2025kce}. For each event, five independent control configurations are generated. In each realization, the $\eta$ and $N$ are preserved, while $\varphi$ of every track is replaced by a value sampled uniformly from the interval $[0,2\pi)$. The normalized Betti curves are computed for each of the configurations and averaged to form a baseline, denoted as $\beta_k^{rand} (\varepsilon)$. This baseline, illustrated in the top panels of Figure \ref{fig6}, quantifies the expected topological evolution of an uncorrelated system with identical longitudinal kinematics and multiplicity. The purely dynamical spatial correlations are then isolated by defining the $\Delta$-Betti curves:
\begin{equation}
\begin{split}
\beta_0^\Delta(\varepsilon) &= \beta_0(\varepsilon) - \beta_0^{\mathrm{rand}}(\varepsilon), \\ \beta_1^\Delta(\varepsilon) &= \beta_1(\varepsilon) - \beta_1^{\mathrm{rand}}(\varepsilon).
\end{split}
\end{equation}
The resulting dynamical topological correlations are isolated in the delta Betti curves, $\beta_k^{\Delta} (\varepsilon)$, shown in the bottom panels of Figure \ref{fig6}. The underlying CMC Lévy walk produces dense spatial configurations that merge and immediately connect at very small $\varepsilon$. In the $\varphi$ randomized version of the same event, those are spread uniformly and thus more components. It means that the unmodified signal event possesses fewer independent connected components than its randomized counterpart at early filtration stages. This results in $\beta_0^{\Delta} (\varepsilon)$ exhibiting a distinct negative dip at small $\varepsilon$ shown in Figure \ref{fig6} for events containing an embedded critical signal. Conversely, the uniform  background of pure EPOS events remains largely invariant under azimuthal randomisation, yielding $\beta_0^\Delta(\varepsilon) \approx 0$. This negative dip serves as the defining topological fingerprint of the critical fluctuations. The subsequent positive peak arises when the randomised event begins to merge into a single giant component, while the spatially separated CMC clusters remain distinct correspondingly, $\beta_1^\Delta(\varepsilon)$ captures the excess of closed loops formed by tightly triangulated CMC clusters. At small $\varepsilon$, the dense CMC clusters solidify into compact filled regions with no internal voids, suppressing the formation of loops, whereas the randomised event's uniform distribution has many transient loops around small voids. This difference manifests as a negative peak. A positive peak near $\varepsilon \approx 0.06$ captures the enclosure of inter-cluster voids as the filtration radius bridges the gaps between separate CMC clusters. For EPOS background (red in Figure \ref{fig6}), both $\beta_0^\Delta$ and $\beta_1^\Delta$ remain almost consistent with zero throughout, confirming that azimuthal randomisation leaves a spatially uniform distribution invariant.

As input to the machine learning classifiers, the $\Delta$-Betti curves are evaluated across $300$ discrete filtration steps. The resulting sequences are concatenated to form a $600$-dimensional feature vector, $\beta_0^\Delta \oplus \beta_1^\Delta$, which encodes the multiscale topological structure of each collision event.

\section{\label{sec:section5}Machine Learning Classification of Topological Features}
The identification of the embedded critical signal (EPOS+CMC) is formulated as a supervised binary classification task. The objective is to discriminate pure EPOS background events (assigned label 0) from the dilute signal mixture containing the CMC injection (assigned label 1). Because the analysis uses most central events only, the resulting classes are naturally balanced, comprising approximately $15 \times 10^3$ events per category. The dataset is partitioned into an 80\% training set and a 20\% independent testing set, with the latter strictly withheld during the training phase to evaluate final model generalization. To guarantee exact reproducibility of the weight initialisation and data partitioning all computational random seeds are fixed.

The $600$-dimensional topological feature vector, generated via the concatenation $\beta_0^\Delta \oplus \beta_1^\Delta$, serves as the exclusive input for classification. To evaluate these features, we deploy two fundamentally distinct machine learning architectures, a one-dimensional convolutional neural network (TopoPointNet)~\cite{Wang:2024bzy} and a classical statistical ensemble of gradient boosted decision trees (XGBoost)~\cite{Chen:2016btl}. Deploying two parallel architectures addresses a critical methodological requirement. Achieving consistent classification performance across both a deep learning framework and a decision tree ensemble validates that the physical discriminating power fundamentally resides within the multiparticle topological features themselves, rather than emerging as a spurious artifact of a specific algorithmic structure. The input representations, layer configurations and primary hyperparameters for both models are summarized in Table \ref{tab:ml_summary}.

\begin{table}[b] 
\caption{\label{tab:ml_summary}Summary of the TPN and BDT architectures.} 
\begin{ruledtabular} 
\begin{tabular}{lcc}
 & TopoPointNet (TPN) & XGBoost (BDT)\\\hline
Input format & $2\times300$ sequence & 600-dim.\ flattened vector \\
Architecture & 1D CNN + Dense & Gradient-boosted trees \\
Conv.\ layers & $2\to128\to256$ (kernel 5) & --- \\
Dense layers & $256\to128\to2$ & --- \\
Trees / estimators & --- & 100 \\
Max depth & --- & 3 \\
Learning rate & $5\times10^{-4}$ (Adam) & 0.05 \\
LR schedule & Cosine annealing & --- \\
Batch size & 32 & --- \\
Dropout & 0.3 & --- \\
Subsample & --- & 0.8 \\
Column sample & --- & 0.8 \\
Regularisation & Batch normalisation & L2 (default) \\
Loss function & Cross-entropy & Log loss \\
Early stopping & Patience = 7 & --- \\
Max epochs & 50 & --- \\
\end{tabular} 
\end{ruledtabular} 
\end{table}

\subsubsection{\label{sec:section5.1}TopoPointNet (TPN)}
The delta Betti curves $\beta_0^\Delta (\varepsilon)$ and $\beta_1^\Delta (\varepsilon)$ constitute ordered sequences of topological information, tracking the structural evolution of the event across the spatial filtration scale. By treating the discrete scale parameter $\varepsilon$ as an ordered sequence, the topological features are naturally processed by a one-dimensional convolutional neural network (1D CNN)~\cite{Huang:2021iux}. TopoPointNet architecture as described in~\cite{Wang:2024bzy} is used in this work to extract localized geometric patterns from these continuous filtration curves. The model ingests the topological data as a two-channel array of length $300$. The initial processing stage consists of a 1D convolutional layer mapping the two input channels to 128 output feature channels, utilizing a local kernel size of 5. This layer detects elementary localized patterns in the topological evolution, such as the initial rapid descent in $\beta_0^\Delta$ indicative of early cluster merging. The convolution is immediately followed by batch normalization to stabilize internal covariate shifts during training and a rectified linear unit (ReLU) activation function to introduce nonlinearity. A second 1D convolutional layer subsequently expands the representation from 128 to 256 channels, again employing a kernel size of 5, batch normalization and a ReLU activation. This deeper layer captures higher-order structural motifs, such as the simultaneous incidence of connected component merging and localized loop formation that characterize the nested geometry of the critical clusters. Following the convolutional layers, a global adaptive max pooling operation reduces the sequence dimension to unity, collapsing the $300$-step spatial sequence into a single 256-dimensional feature vector. This max-pooling mechanism imposes translation invariance along the filtration axis. It guarantees that the network registers the presence of a critical topological signature irrespective of the absolute scale $\varepsilon$ at which the structures become active in the Delaunay sub-level-set filtration. The network is therefore sensitive exclusively to the intrinsic correlational geometry of the embedded signal, rather than rigidly anchoring to specific absolute inter-particle distances.

\begin{figure*}[htb]
    \centering
    \includegraphics[width=0.48\textwidth]{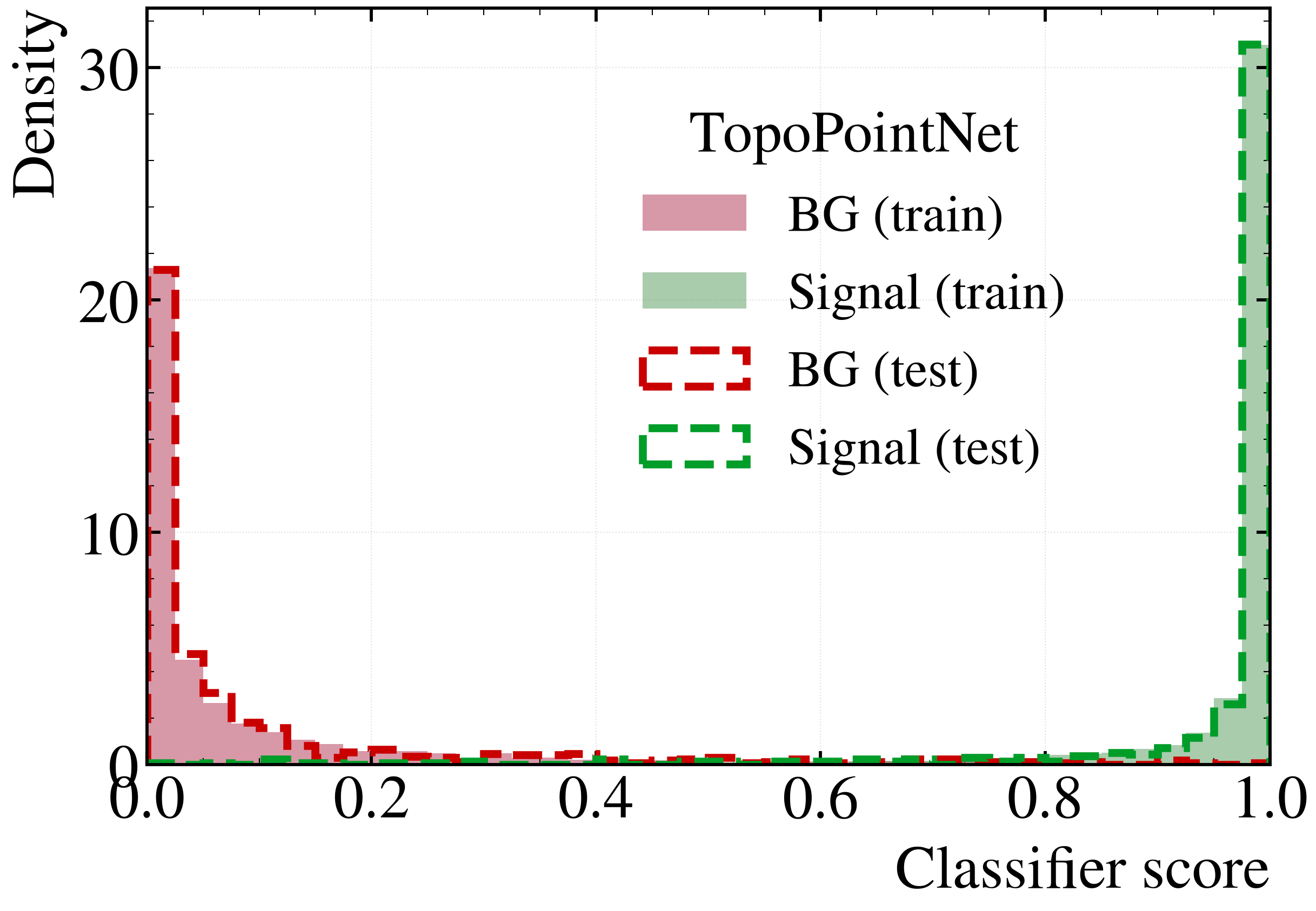}
    \includegraphics[width=0.48\textwidth]{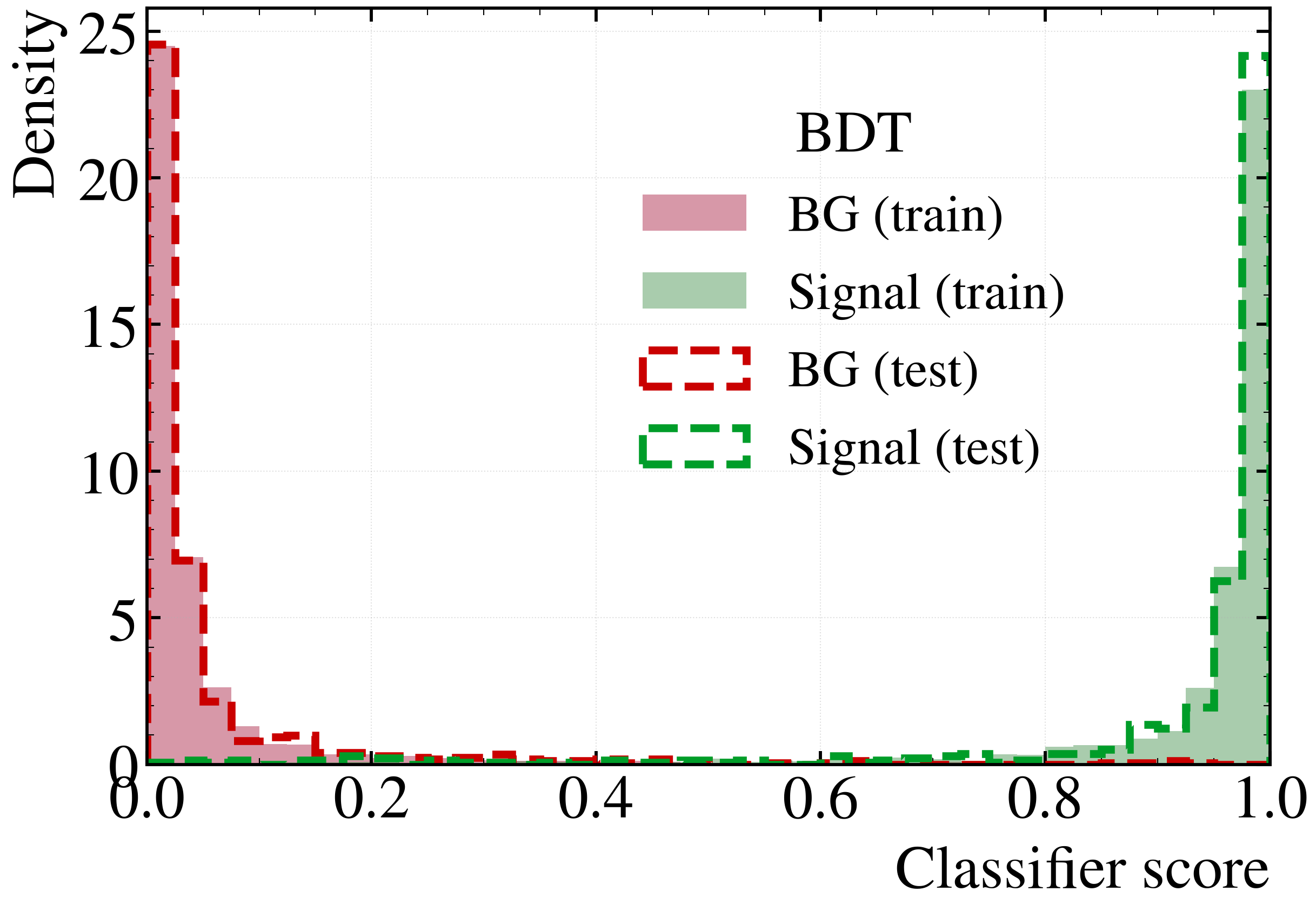}
    \caption{\label{fig7a}Output classifier score distributions $P(S_{\mathrm{ML}})$ for TopoPointNet (left panel) and BDT (right panel) evaluated on training and test events samples. The x-axis denotes the continuous classifier score $S_{\mathrm{ML}} \in [0, 1]$ (signal probability) and the y-axis represents the probability density $P(S_{\mathrm{ML}})$ (normalized unit area). Filled (red) area represents EPOS background training events (peaking near $S_{\mathrm{ML}} \to 0$), green filled represents 5\% CMC signal-injected training events (skewing toward $S_{\mathrm{ML}} \to 1$). Dashed steps correspond to the independent testing event sample for background (red) and signal (green). The strict overlap between training and testing distributions confirms the absence of model overtraining. }
\end{figure*}

\subsubsection{\label{sec:section5.2}Boosted Decision Trees (BDT)}
As an alternative architecture to the deep convolutional network, the eXtreme Gradient Boosting (XGBoost) algorithm~\cite{Chen:2016btl} is used. The foundational element of this model is the decision tree, which partitions the topological feature space into discrete regions through a sequence of binary threshold splits, assigning a constant predicted score to each terminal leaf. Because a single shallow tree lacks the capacity to capture complex multiparticle correlations, the boosting algorithm constructs an additive ensemble sequentially. Each successive tree is trained to fit the residuals of the preceding ensemble's predictions, thereby executing gradient descent in the functional space. XGBoost specifically optimizes this sequential learning by expanding the arbitrary differentiable loss function to second order. At iteration $t$, the algorithm minimizes the approximated objective, \begin{equation} \tilde{L}^{(t)} = \sum_{i} \left[ g_i f_t(x_i) + \frac{1}{2} h_i f_t^2(x_i) \right] + \Omega(f_t), \end{equation} where $f_t(x_i)$ represents the prediction of the newly added tree for the topological feature vector $x_i$ of the $i$-th event, $g_i$ and $h_i$ are the respective first and second-order gradient statistics of the loss function and $\Omega(f_t)$ explicitly penalizes tree complexity. This second-order formulation allows the classifier to rapidly converge on the optimal topological decision boundaries without over-fitting the limited training sample~\cite{Chen:2016btl}.

To interpret the topological data with BDT, the $600$-dimensional delta Betti feature sequence $\beta_0^\Delta \oplus \beta_1^\Delta$ is supplied as a flattened, 1-D array. While the 1-D convolutional network extracts hierarchical motifs from the ordered filtration steps, the decision tree ensemble evaluates the topological invariants at each discrete scale parameter $\varepsilon$ as independent scalar features. To physically interpret the classification boundaries constructed by the tree ensemble, SHapley Additive exPlanations (SHAP)~\cite{lundberg:2017} is utilized in this work. SHAP provides a unified measure of feature attribution derived from cooperative game theory. It calculates the expected marginal contribution of each input feature to the final model prediction by averaging over all possible feature permutations. This assigns an exact attribution weight to each of the $600$ discrete filtration steps, revealing exactly which spatial scales drive the model's capacity to identify critical events. To validate the SHAP attributions, the inherent gain based feature importance of the XGBoost is computed in parallel. This internal metric quantifies the fractional reduction in the objective function achieved by tree splits utilizing a given topological feature. Extracting these importance metrics across the filtration axis isolates the spatial scales containing the primary physical discriminating power. As will be quantified in Sec.~\ref{sec:section6.2}, both the SHAP attributions and the gain-based importance exhibit a strict peak at filtration scales of $\varepsilon \approx 0.02–0.10$. The convergence of these interpretability metrics demonstrates that the model does not rely on random statistical artifacts, rather it independently identifies the exact characteristic angular scale at which the nested geometry of the critical clusters diverges from the thermal background.

\begin{figure*}[htb]
    \centering
    \includegraphics[width=0.32\textwidth]{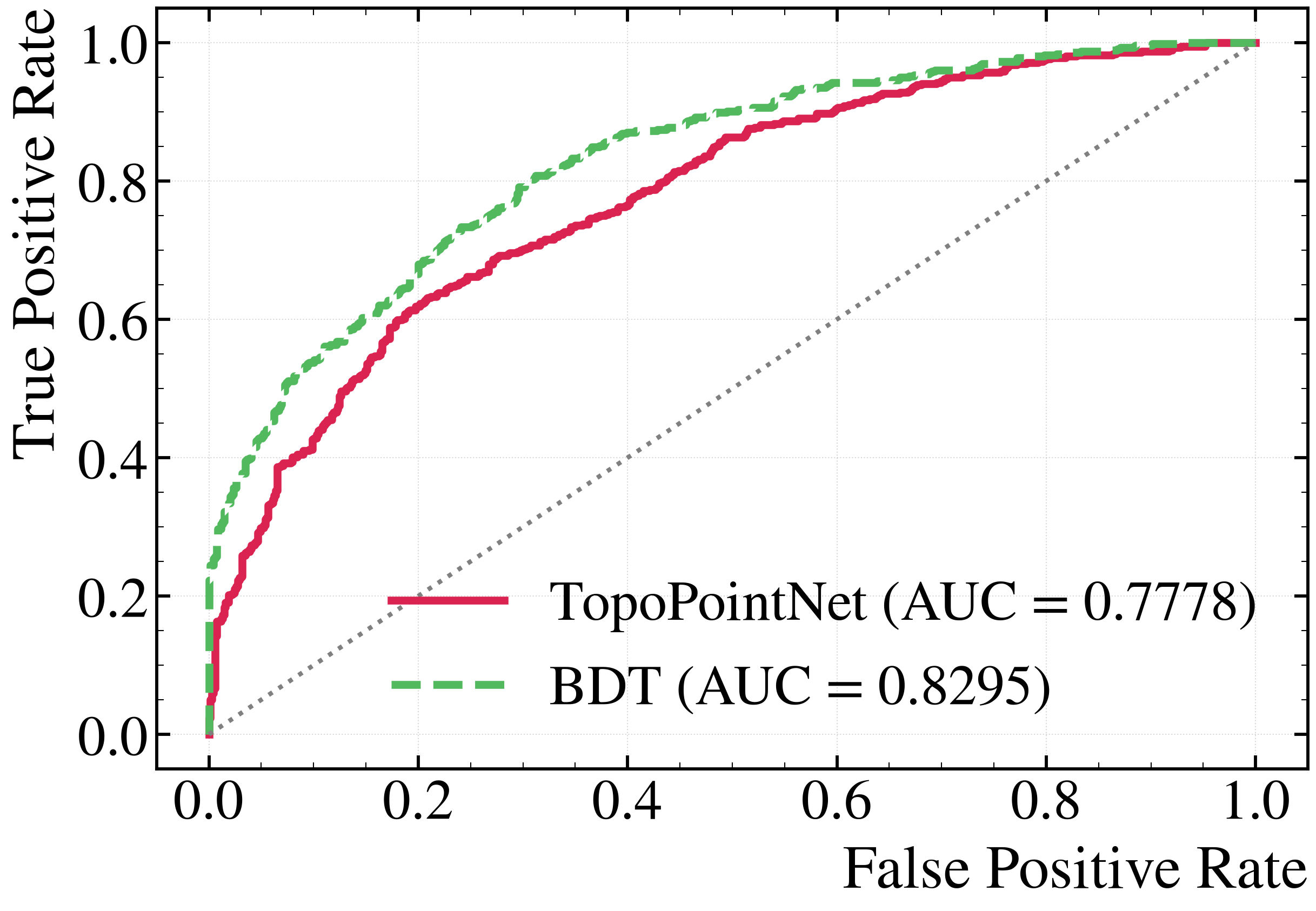}
    \includegraphics[width=0.32\textwidth]{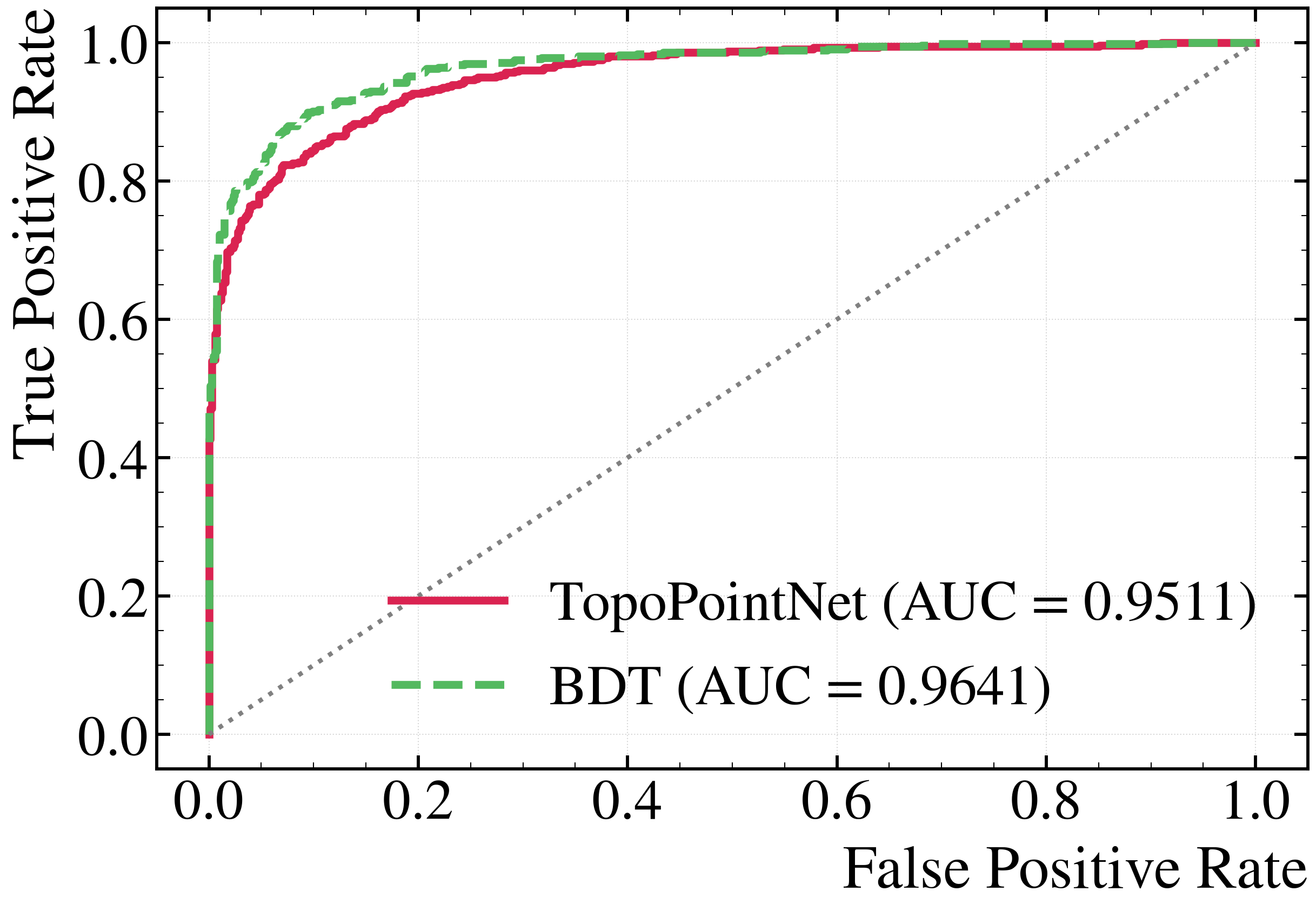}
    \includegraphics[width=0.32\textwidth]{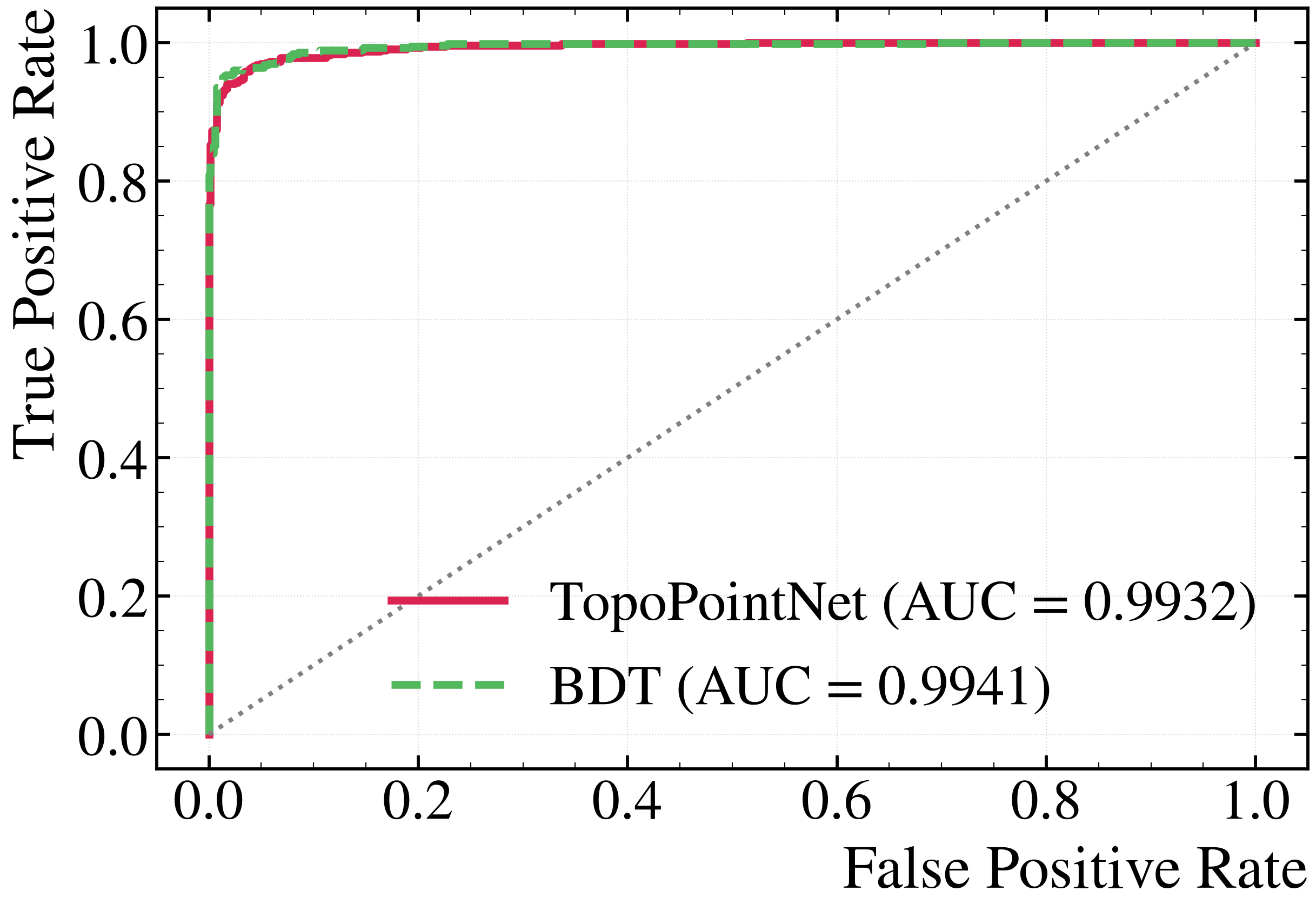}
    \caption{\label{fig7b}ROC performance curves for TopoPointNet (red) and BDT (green) across signal replacement fractions 1\% (left, $\text{AUC} \approx 0.78 - 0.83$), 3 \% (middle, $\text{AUC} \approx 0.95-0.96$), and 5\% (right, $\text{AUC} \approx 0.99$).}
\end{figure*}

\section{\label{sec:section6}Results}
The extraction of the intermittency signal from the dense background requires isolating correlations at both the macroscopic event scale and the microscopic track scale. To achieve this a unified two-stage topological filtering scheme is applied. The first stage executes an event-level selection utilizing the continuous classification scores generated by the machine learning architectures. Each event is assigned a signal probability score, $\mathcal{S} \in [0,1]$, defined as the posterior probability output by the classifier's final softmax layer. It is the probability that the event belongs to the signal class rather than the background class.  Events are retained for subsequent analysis if their predicted signal probability satisfies the threshold condition ($\mathcal{S} \geq 0.90$). The criteria filters the event sample to isolate configurations highly enriched with critical fluctuations.
Even within this new sample, the embedded critical particles remain a dilute minority. To separate the signal from the underlying background, the second stage applies a geometric particle-level density filter directly in the $(\eta, \varphi)$ phase space. Within each selected event, individual tracks are retained if their evaluated nearest-neighbour distance satisfies \two{d}{NN} $\leq \varepsilon_{cut}$. This angular cut isolates the tightly packed particles comprising the critical clusters while efficiently discarding the diffuse thermal bulk. The specific threshold value of $\varepsilon_{cut}$ is physically motivated by the feature attribution frameworks.

\begin{figure*}[htb]
    \centering
    \includegraphics[width=0.48\textwidth]{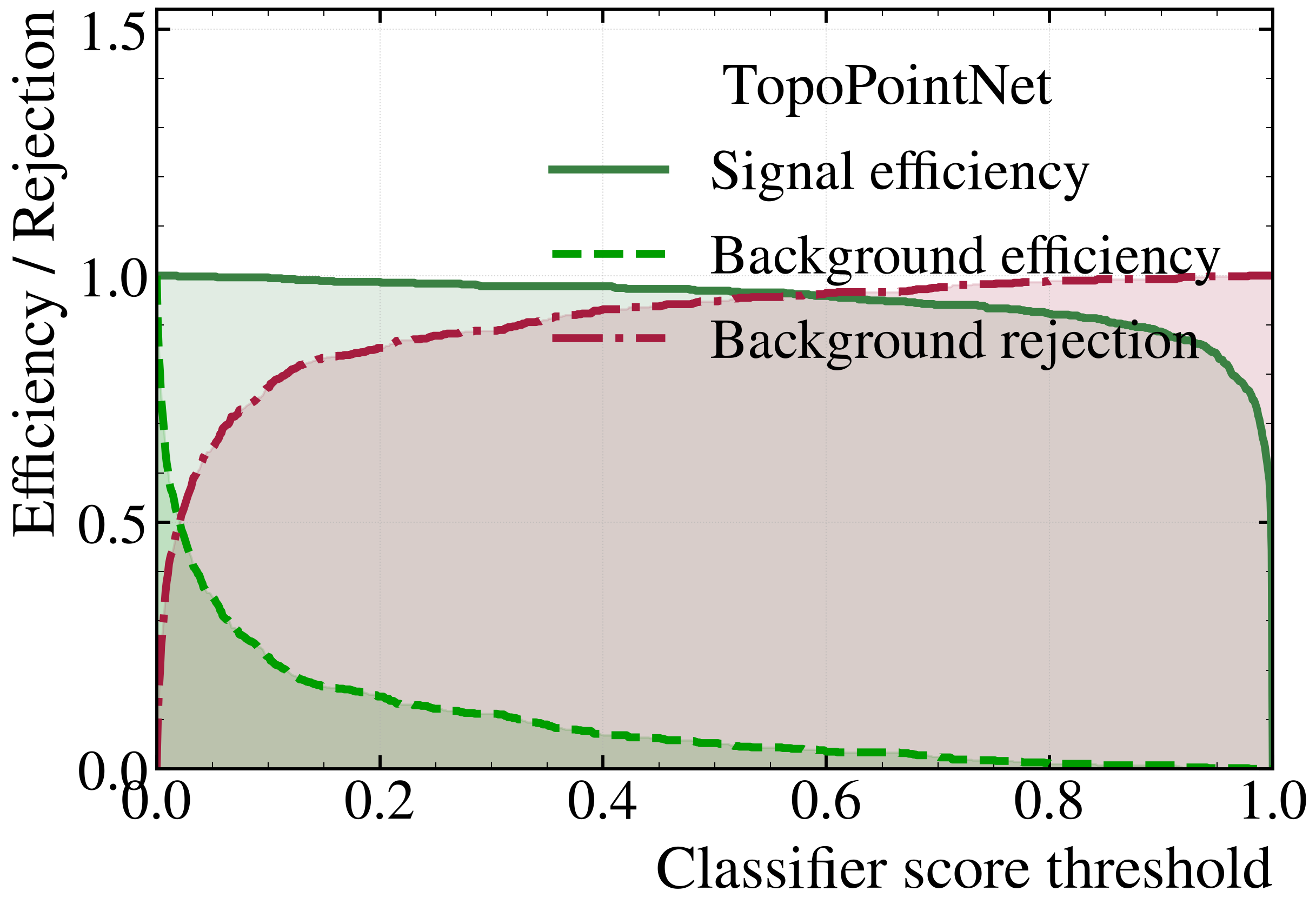}
    \includegraphics[width=0.48\textwidth]{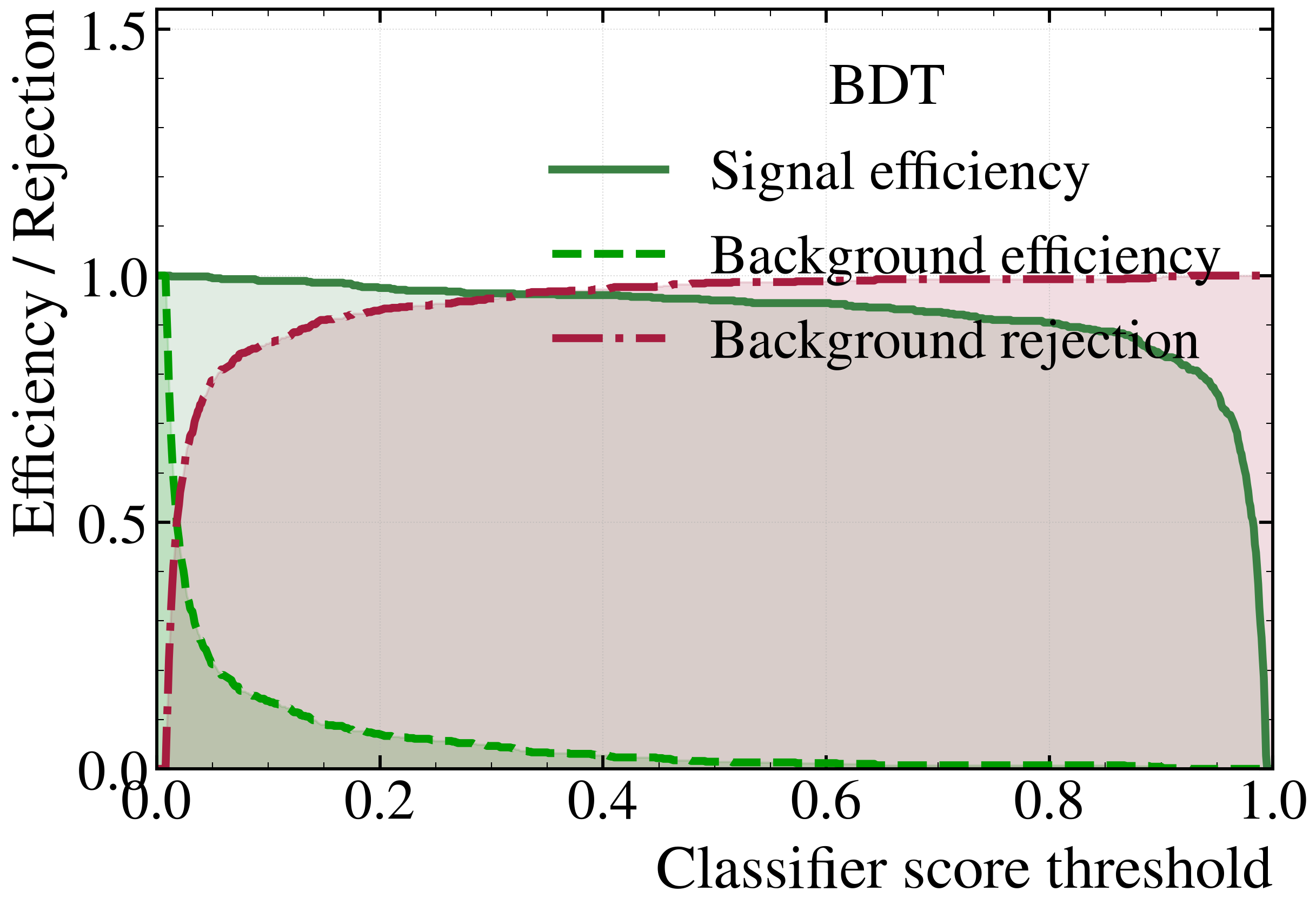}
    \caption{\label{fig8}Signal selection efficiency $\varepsilon_{\mathrm{sig}}$ (green solid), background efficiency $\varepsilon_{\mathrm{bkg}}$ (green dashed) and background rejection fraction $1 - \varepsilon_{\mathrm{bkg}}$ (red dashed) vs ML score threshold for TopoPointNet (left panel) and BDT (right panel) evaluated on test samples of 5\% CMC signal-injected events. y-axis denotes the fraction of events ($[0, 1]$).
    %Classifier performance metrics as a function of the score threshold $S_{\mathrm{thresh}} \in [0, 1]$ for TopoPointNet (left panel) and BDT (right panel), evaluated on test samples of 5\% CMC signal-injected events ($\lambda = 0.05$) vs pure EPOS background in central Pb--Pb collisions at \energy{5.02}. The vertical axis denotes the fraction of events ($[0, 1]$). Dark green solid lines represent signal selection efficiency $\varepsilon_{\mathrm{sig}}$ (fraction of true signal events passing $S_{\mathrm{ML}} \ge S_{\mathrm{thresh}}$); bright green dashed lines represent background efficiency $\varepsilon_{\mathrm{bkg}}$ (fraction of pure thermal EPOS background events falsely passing the threshold); red dash-dotted lines represent background rejection $1 - \varepsilon_{\mathrm{bkg}}$ (fraction of EPOS background events correctly rejected). 
    %At the primary working threshold $S_{\mathrm{thresh}} = 0.90$, background efficiency drops to $\varepsilon_{\mathrm{bkg}} \lesssim 5\%$ (background rejection $> 95\%$), ensuring a signal-enriched event ensemble for Stage 2 particle-level filtering.
    %The vertical dash-dotted line marks the working threshold $S_{\mathrm{thresh}} = 0.90$.
    }
\end{figure*}

\subsubsection{\label{sec:section6.1}Classification performance}
A continuous classification score bounded within the interval is evaluated, representing the likelihood of an event containing the embedded critical signal. Figure~\ref{fig7a} shows the resulting score distributions for signal ($\lambda = 0.05$) and background, normalized to unit area (probability density). For both the TPN and the BDT ensemble, pure EPOS events are mostly concentrated near a score of zero, while the signal mixture events accumulate near one. The score distributions of the training and independent testing datasets exhibit strict overlap, showing that the classifiers do not suffer from statistical overtraining given the limited sample size. Despite the separation at the extremes, a large population still persists between scores of 0.3 and 0.8. This overlap reflects the fundamental physical difficulty of the classification task. At an injection fraction of $\lambda = 0.05$, exactly 95\% of the final-state kinematics in a signal event originate from the background. In events where the localized Lévy walk generates a highly dispersed or extremely low-multiplicity cluster, the overall topological structure remains dominated by the diffuse background, rendering event-level classification intrinsically ambiguous. 

The diagnostic capability of the topological pipeline is highly dependent on the signal injection fraction $\lambda$. To quantify this sensitivity, we evaluate the Receiver Operating Characteristic (ROC) curves across different CMC signal fractions: 1\%, 3\% and 5\%, as shown in Figure~\ref{fig7b}. The Area Under the Curve (AUC) strictly degrades as the critical signal becomes more dilute. At 1\%, the classification performance in Figure \ref{fig7b} (left) is highly constrained, yielding modest AUC values of $0.78$ for TPN and $0.83$ for BDT. As the signal concentration increases to 3\% in Figure \ref{fig7b} (middle), the emergence of more distinct topological signatures enhances the discriminative capability, with the AUC rising to $0.95$ for TPN and $0.96$ for BDT.  At 5\% in Figure \ref{fig7b} (right), the models achieve their highest discrimination power, yielding an AUC of approximately $0.99$ for TPN and $0.99$ for BDT. This performance establishes 5\%$ (\lambda = 0.05$) as an effective threshold for event-level discrimination. The number of discrete filtration steps $N_\varepsilon$ constitutes a resolution hyperparameter governing the granularity at which the Betti curves sample the topological evolution. To assess the sensitivity of the classification performance to this parameter, the full pipeline is evaluated at three resolutions, $N_\varepsilon = 100$, $300$ and $500$. At the baseline resolution of $N_\varepsilon = 100$, the BDT achieves an AUC of $0.954$ and the TPN an AUC of $0.942$ for $\lambda = 0.05$. Increasing the resolution to $N_\varepsilon = 300$ yields a substantial improvement, with the AUC rising to $0.994$ for the BDT and $0.993$ for the TPN. A further increase to $N_\varepsilon = 500$ produces only a marginal gain (AUC $\approx 0.999$ for the BDT and $0.998$ for the TPN), indicating that the classification performance has effectively converged. The enhanced resolution resolves the fine-grained topological structure of the tightly packed Lévy-walk clusters at small $\varepsilon$, where the 100-step grid undersamples the rapid evolution of connected components. Beyond $N_\varepsilon = 300$, the additional filtration steps predominantly sample the featureless thermal bulk at intermediate and large $\varepsilon$, contributing negligible discriminating information. Accordingly, $N_\varepsilon = 300$ is adopted as the operational resolution for all subsequent results.

% Dark green solid lines represent signal selection efficiency $\varepsilon_{\mathrm{sig}}$ (fraction of true signal events passing $S_{\mathrm{ML}} \ge S_{\mathrm{thresh}}$); bright green dashed lines represent background efficiency $\varepsilon_{\mathrm{bkg}}$ (fraction of pure thermal EPOS background events falsely passing the threshold); red dash-dotted lines represent background rejection $1 - \varepsilon_{\mathrm{bkg}}$ (fraction of EPOS background events correctly rejected). 

Figure~\ref{fig8} shows the classification performance quantified by the signal efficiency $\varepsilon_{\mathrm{sig}}$ and background rejection $1 - \varepsilon_{\mathrm{bkg}}$ as a function of the classifier score threshold. Both architectures exhibit strong separation over the full threshold range. At a loose threshold of $\mathcal{S} \geq 0.50$, the TPN retains 96.9\% of signal events while rejecting 94.8\% of the background, the BDT retains 95.0\% of signal events while rejecting 98.5\%. The equal-efficiency crossover point, where $\varepsilon_{\mathrm{sig}} = 1 - \varepsilon_{\mathrm{bkg}}$, occurs at $\mathcal{S} \approx 0.59$ for TPN and $\mathcal{S} \approx 0.33$ for BDT, with both models achieving a balanced operating value of approximately 96\% at their respective crossover thresholds. For the subsequent track-level filtration, the event-level selection is fixed at the strict operating point of $\mathcal{S} \geq 0.90$. At this threshold, the TPN retains 88.7\% of the true signal events while rejecting 99.3\% of the pure thermal EPOS background, the BDT retains 84.0\% of the signal while rejecting 99.6\%. This conservative operating point prioritises sample purity over efficiency, yielding a strongly signal-enriched event sample for the second filtering stage.

The performance metrics from TPN and the BDT ensemble agree closely. This consistency across two independent architectures indicates that the discriminating power resides in the multiscale topological geometry of the point clouds themselves, not in the choice of a particular machine learning model. The BDT is thus adopted as the primary interpretable model for the remainder of the analysis.
%, owing to its exact, tree-based SHAP attribution that permits direct identification of the discriminating filtration scales without the approximations required by gradient-based attribution methods for neural networks.

\subsubsection{\label{sec:section6.2}Classifier feature attribution}

\begin{figure*}[htb]
    \centering
    \includegraphics[width=0.48\textwidth]{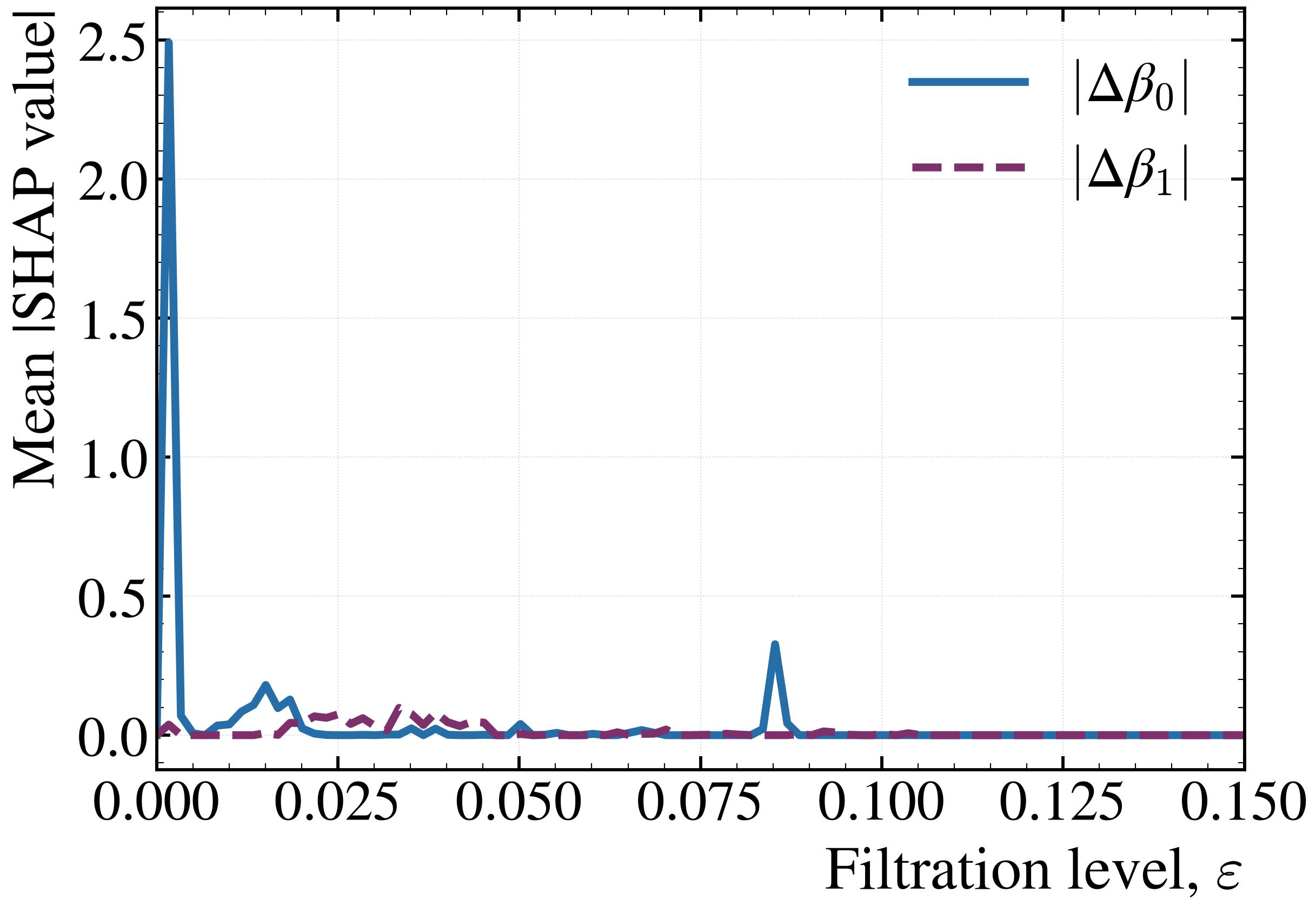}
    \includegraphics[width=0.48\textwidth]{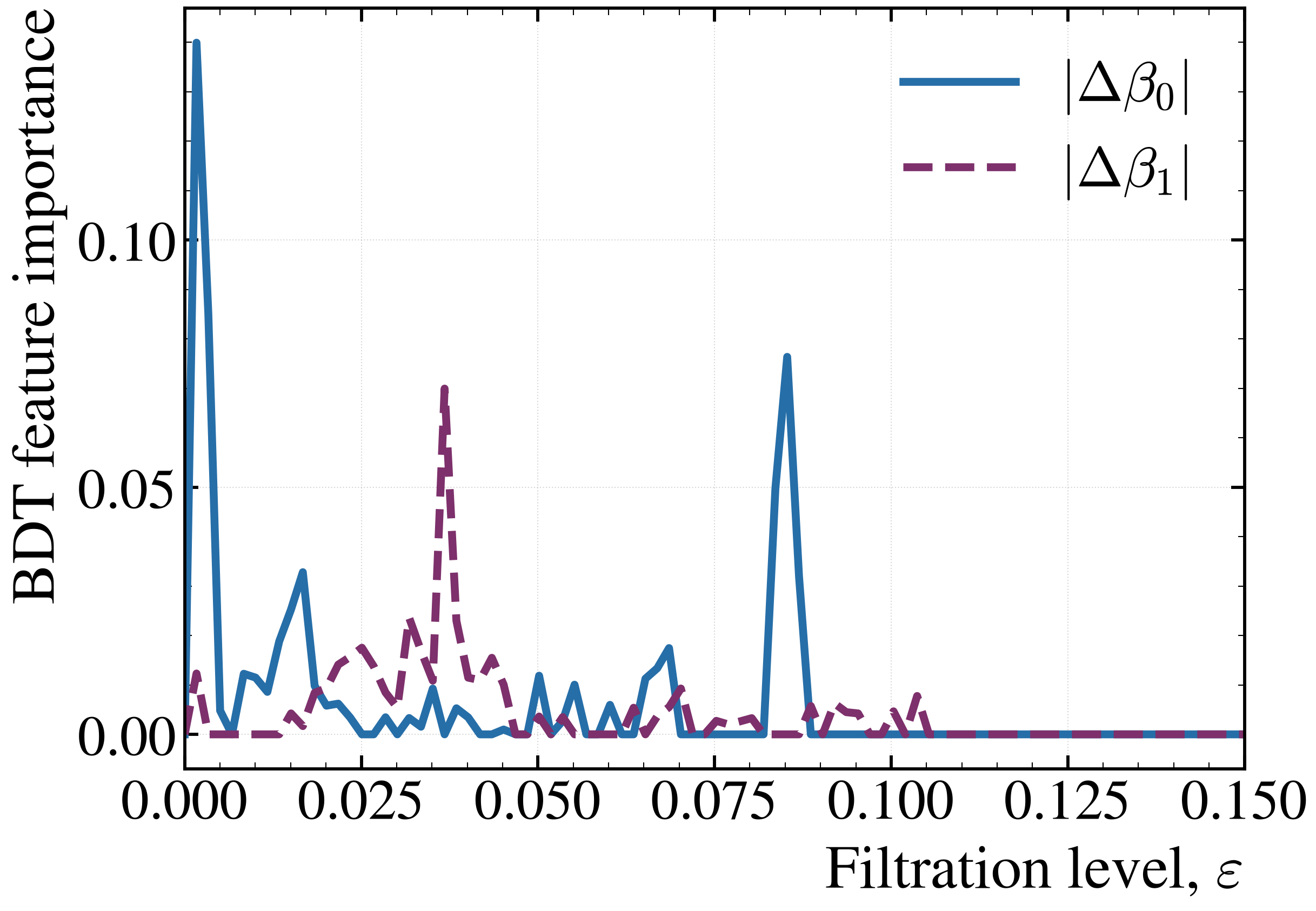}
    \caption{\label{fig9} Topological feature importance across Delaunay filtration scale, $\varepsilon$ for the BDT model on $\Delta$-Betti features ($\Delta\beta_0$, blue lines; $\Delta\beta_1$, red lines). Mean absolute SHAP (left panel) values quantifying the additive contribution of each filtration scale $\varepsilon$ to the model predictions. Gain-based BDT feature importance (right panel) across tree splits, which exhibits a secondary peaks corresponding to the cluster-coalescence regime where distinct Levy clusters merge into macroscopic topological structures.}
\end{figure*}

To interpret the physical mechanisms driving the BDT classification decisions, the global feature importance is evaluated as a function of the filtration parameter $\varepsilon$. Two complementary interpretability metrics are employed, the mean absolute Shapley value~\cite{lundberg:2017} averaged across the test ensemble for each discrete $\Delta$-Betti feature (SHAP) and the intrinsic gain-based feature importance, measuring the average fractional reduction in the training loss contributed by tree splits acting on each topological variable~\cite{Chen:2016btl}.

As shown in Fig.~\ref{fig9}, the two BDT interpretability metrics reveal a physically rich, scale-resolved picture. The SHAP attribution is overwhelmingly dominated by a sharp primary peak at $\varepsilon \approx 0.001$ rad in the $|\Delta\beta_0|$ channel, with a mean absolute SHAP value of approximately 3.5, exceeding all other features by more than an order of magnitude. This filtration scale corresponds to the smallest nearest-neighbour distances in the Delaunay complex, at which the most tightly packed particles within the Lévy-walk clusters first activate as connected components. A secondary, lower-amplitude peak appears in the $|\Delta\beta_1|$ channel, distributed across $\varepsilon \approx 0.02$--$0.05$ rad. This scale corresponds to the regime in which 1-cycles (closed triangulated loops) form within the signal topology as intra-cluster triangulations solidify. A third, isolated peak in $|\Delta\beta_0|$ emerges at $\varepsilon \approx 0.084$ rad, attributable to the cluster-coalescence regime in which spatially separated CMC clusters begin to merge into macroscopic connected components. The gain-based importance exhibits a consistent hierarchical structure, with the dominant peak again concentrated at $\varepsilon \approx 0.001$ rad in $|\Delta\beta_0|$, confirming that the earliest tree splits exploit the same ultra-small-scale density contrast identified by SHAP. The secondary gain peaks are broadly distributed across $\varepsilon \approx 0.02$--$0.05$ and $\varepsilon \approx 0.08$--$0.10$ rad, mirroring the loop-formation and coalescence regimes identified above. The concordance between the two independent attribution frameworks confirms that the BDT derives its primary discriminating power from the onset of connectivity at the characteristic angular scale of the critical clusters.

\subsubsection{\label{sec:section6.3}Recovery of the NFMs}

\begin{figure}[h!]
    \includegraphics[width=0.49\textwidth]{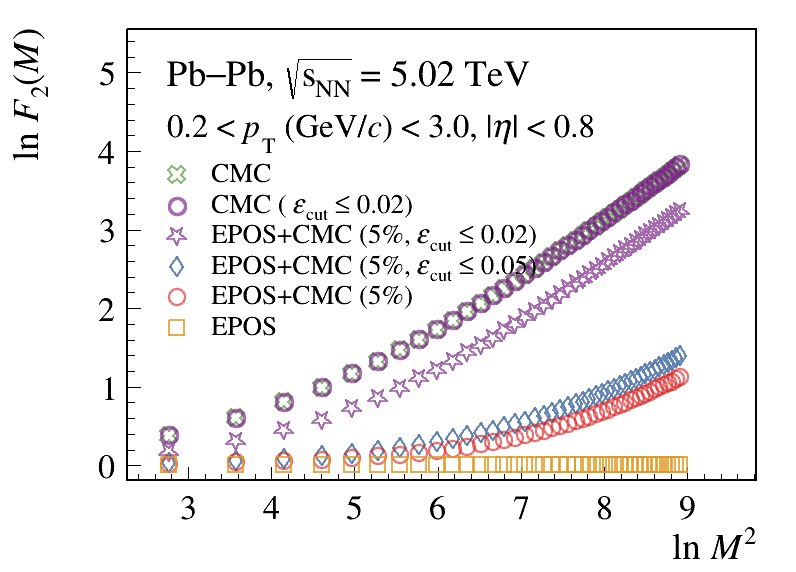}
    \caption{\label{fig10} Restoration of intermittency scaling, \fq{2}  vs $\ln M^2$ using the two-stage topological ML filtration  in central Pb--Pb collisions at \energy{5.02}. The recovered trend and slope (magenta stars) with $\varepsilon_{\mathrm{cut}} \leq 0.02\text{ rad}$ restores the pure CMC (green crosses) from the 5\% diluted mixture (red circles). Filtered pure CMC (magenta circles) are shown to verify that the density cut does not distort the intrinsic critical scaling. A looser cut $\varepsilon_{\mathrm{cut}} \le 0.05\text{ rad}$ (blue diamonds) is also included to show the gradual effect of $\varepsilon_{\mathrm{cut}}$.}
\end{figure}

\begin{figure}[h!]
    \includegraphics[width=0.49\textwidth]{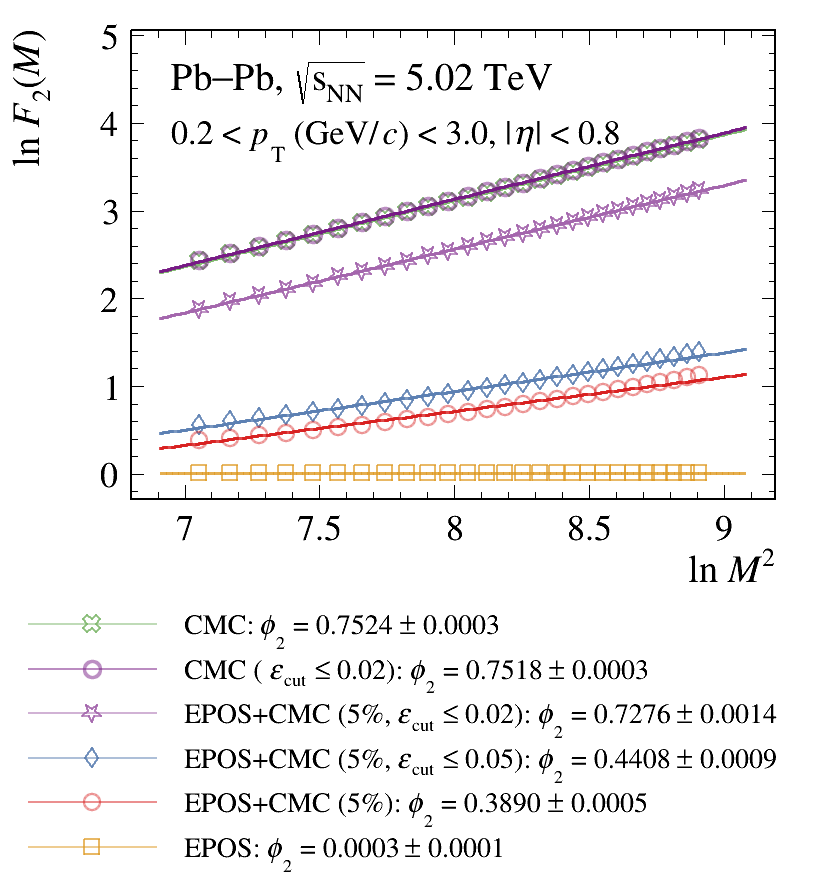}
    \caption{\label{fig11} Recovery of intermittency index, \two{\phi}{2} after two-stage topological ML filtration in central Pb--Pb collisions at \energy{5.02}. Linear fits to calculate \two{\phi}{2} are performed in higher $M^2$ region ($\in [6.9, 9.1]$). Results compare pure CMC (green crosses), 5\% signal mixture (red circles), filtered 5\% signal mixture (magenta stars) and EPOS background (yellow squares). The optimal cut recovers $\phi_2 \approx 0.73$, restoring the pure CMC benchmark ($\phi_2 \approx 0.75$) from the diluted sample.}
\end{figure}

To quantify the recovery of the critical signal, the benchmarks by computing have been established in Figure~\ref{fig2a},\ref{fig3}. Applying the primary stage of the topological pipeline filters the event ensemble, retaining only those collision configurations that satisfy the $\mathcal{S}\geq0.90$ threshold. While this event-level classification successfully isolates a highly purified sample enriched with critical fluctuations, computing \two{F}{q}($M$) on these retained events yields a nearly flat scaling behaviour. Because the injection fraction is fixed at a $\lambda$ value, the remaining final-state tracks within any correctly identified signal event still originate from the uniform thermal background.
This severe spatial dilution within the selected events demonstrates that macroscopic event-level discrimination is fundamentally insufficient to restore the critical scaling. The underlying power-law behavior cannot be extracted without explicitly stripping the diffuse thermal tracks from the signal configurations, dictating the necessity of the microscopic, particle level density filter.

To overcome the background domination, the second stage applies the particle level density filter to the event sample. Within these signal enriched events, individual tracks are retained exclusively if their local nearest-neighbour distance satisfies \two{d}{NN}$\leq$ 0.02 rad. This geometric threshold explicitly strips away the diffuse thermal bulk, isolating the densely packed particles comprising the embedded critical clusters. Re-evaluating the NFMs on this filtered track sample yields a clear restoration of the power-law scaling. Fig.~\ref{fig10} demonstrate the scaling behaviour of ln \two{F}{2}($M$) as a function of ln $M^2$ across the EPOS background, the pure CMC reference and the unfiltered 5\% signal mixture, along with the fully filtered pipeline samples at different filtration thresholds, \two{\varepsilon}{cut}. Among these, the \two{\varepsilon}{cut} $\geq 0.02$ sample lies closest to the pure CMC reference, where a looser cut of 0.05 retains an excess of background tracks. To verify that the selection does not itself distort the intrinsic scaling of the critical signal, the cuts are applied to the pure CMC sample. The filtered CMC scaling is found to be indistinguishable from the unfiltered case, confirming that the filtration preserves the intermittency signal. 

The linear fit to the double-logarithmic distribution of the samples is compiled in Figure~\ref{fig11}. The recovered intermittency index of $\phi_2 \approx 0.73$ for the filtered 5\% signal mixture sample, in close agreement with the pure CMC reference of $\phi_2 \approx 0.75$. A residual discrepancy of approximately 1\% persists between the recovered indices and the pure reference values. This minor suppression is attributable to surviving contamination from dense EPOS background fluctuations that accidentally satisfy the tight spatial threshold, compounded by the finite signal efficiency of the initial event-level classification stage. The two-stage topological methodology thus successfully disentangles the critical geometry from a highly dilute sample, allowing for a precise recovery of the underlying intermittency scaling.

\section{\label{sec:section7}Conclusions and outlook}

The search for signatures of the QCD phase transition and critical point via scale-invariant multiplicity fluctuations requires isolating extremely dilute correlation signals from an overwhelming non-critical background. In this work, the first implementation of the Critical Monte Carlo (CMC) model within the two-dimensional angular $(\eta, \varphi)$ phase space is presented, establishing the critical geometry for Pb--Pb collisions at \energy{5.02} simulated with the EPOS event generator. The pure CMC reference yields a second-order intermittency index of $\phi_2 \approx 0.75$ in this geometry, providing a shifted baseline relative to the theoretically predicted value of $\phi_2 = 2/3$ in $(p_x, p_y)$ space, attributable to the coordinate transformation and finite-size acceptance boundaries. Embedding the CMC signal at experimentally realistic fractions ($\lambda = 0.01$--$0.05$) into the EPOS background demonstrates severe signal dilution. The resulting $\phi_2$ values collapse toward the inert background, confirming that standard factorial moment analysis is fundamentally insufficient to extract weak critical fluctuations from a dominant background. To overcome this dilution, a two-stage topological machine learning pipeline was constructed. Final-state particle distributions are represented as two-dimensional point clouds in $(\eta, \varphi)$ space, upon which a periodic Delaunay triangulation and a nearest-neighbour sub-level-set filtration are constructed across $N_\varepsilon = 300$ discrete scale steps. Raw Betti curves $\beta_0(\varepsilon)$ and $\beta_1(\varepsilon)$ are corrected for trivial multiplicity bias via event-wise azimuthal randomisation, yielding $\Delta$-Betti curves that encode exclusively the dynamical multi-particle correlations. The resulting $600$-dimensional feature vectors serve as input to two parallel and architecturally independent classifiers, a one-dimensional convolutional TopoPointNet and a gradient-boosted decision tree (BDT) ensemble. The filtration resolution $N_\varepsilon$ was systematically optimised by evaluating the classification performance at $N_\varepsilon = 100$, $300$ and $500$, the AUC converges beyond $N_\varepsilon = 300$, which is adopted as the operational resolution. Both architectures achieve consistent and high discrimination power at a 5\% signal injection fraction, yielding AUC values of $0.993$ (TPN) and $0.994$ (BDT). This architectural independence confirms that the classification is strictly driven by the underlying topological geometry of the point clouds rather than by algorithmic properties. At the operating threshold of $\mathcal{S} \geq 0.90$, the TPN retains $88.7\%$ of the true signal events while rejecting $99.3\%$ of the pure thermal background; the BDT retains $84.0\%$ of the signal while rejecting $99.6\%$, yielding a strongly signal-enriched event ensemble for the second filtering stage. Feature attribution via SHAP and gain-based importance identifies the dominant discriminating scale at $\varepsilon \approx 0.001$ rad in the $|\Delta\beta_0|$ channel, corresponding to the characteristic nearest-neighbour distance of the tightly packed Lévy-walk clusters. Secondary peaks in the $|\Delta\beta_1|$ channel at $\varepsilon \approx 0.02$--$0.05$ rad mark the loop-formation regime, while a tertiary peak at $\varepsilon \approx 0.08$ rad corresponds to the cluster-coalescence scale. The concordance of the two independent attribution frameworks validates the physical interpretability of the classification boundaries. While event-level classification alone successfully purifies the event sample, the NFMs evaluated on these selected events remain flat, as the embedded critical particles constitute a dilute minority of the accepted tracks. The secondary particle-level density filter, applied at $\varepsilon_{\mathrm{cut}} = 0.02$ rad, strips the diffuse thermal bulk and isolates the densely packed critical clusters. The combined two-stage filtration successfully restores the power-law scaling of the factorial moments, recovering an intermittency index of $\phi_2 \approx 0.73$ from the 5\% diluted mixture, in close agreement with the pure CMC reference value of $\phi_2 \approx 0.75$. The density cut applied to the pure CMC sample produces no distortion of the intrinsic scaling, confirming that the filtration preserves the critical geometry. These results establish topological machine learning as a robust, data-driven methodology for probing the QCD critical point and the phase structure of strongly interacting matter through scale-invariant spatial fluctuations in heavy-ion collisions at LHC energies.

While the topological pipeline successfully recovers the underlying critical geometry, several methodological and experimental limitations must be addressed prior to its deployment on real collision data. First, the present analysis is conducted strictly at the particle generator level. Experimental realities, such as finite track momentum resolution and track merging or splitting artifacts, are not simulated. In modern collider experiments, tracks separated by angular distances of close $\eta$ or $\varphi$ can often merge into a single reconstructed track~\cite{ALICE:2014sbx}. Because the second-stage geometric filter relies on a nearest-neighbour threshold, it operates precisely within this sensitive detector resolution regime. A comprehensive simulation of the specific detector response is therefore necessary to validate the survival of the topological signatures against tracking inefficiencies. Second, while the event level classification and the particle level filtration operate sequentially, they are not strictly statistically independent. The deep learning and decision tree architectures evaluate multiscale Betti curves derived directly from the local spatial sub-level-set filtration, which is governed by the identical \two{d}{NN} metric used to strip the background tracks.

The experimental deployment relies on the enhanced data acquisition capabilities of the ALICE experiment during the LHC Run 3. The upgraded continuous readout of the Time Projection Chamber sampled an integrated luminosity of 10$nb^{-1}$ for Pb--Pb collisions~\cite{Lippmann:2014lay}, delivering the massive statistical precision necessary to evaluate highly differential multiscale Betti curves on an event-by-event basis.

While the present topological filtration is constructed exclusively within the angular $(\eta, \varphi)$ acceptance, the persistent homology framework mathematically extends to ($p_x, p_y$) as well. However, projecting the topological features into the non-uniform grid introduces geometric complications. 
%Because $\eta$ has smaller units than $\varphi
The kinematic variables must thus be mapped onto a uniform domain via cumulative transformations~\cite{Bialas:1990dk,DeWolf:1995nyp}. Applying the Delaunay sub-level-set filtration within this flattened cumulative coordinate space guarantees that the resulting simplicial complexes reflect purely dynamical correlation structures rather than the trivial single-particle distributions.

Finally, the CMC algorithm provides a static, explicitly scale-invariant fractal geometry. To determine the sensitivity of the topological machine learning architectures to dynamic phase evolution, future calibrations will replace the static CMC injection with configurations generated by the Successive Contraction and Randomization (SCR) model~\cite{Hwa:2011bu}. The SCR framework mimics the quark-hadron phase transition by iteratively applying localized spatial contractions representing confinement attraction, counterbalanced by continuous thermal randomization. Extracting the $\Delta$-Betti feature sequences from these dynamically evolved events will test the capacity of the pipeline to recover critical intermittency indices from systems where the multiscale geometry emerges organically from the interplay of opposing physical forces.

\begin{acknowledgments}
We sincerely thank the authors of EPOS. We gratefully acknowledge the Research and Seed Grant under the Quality Assurance Fund (DIQA), University of Jammu (Sanction No.~RA/23/1293-1300dl-7/8/2023), for the financial support for the computing facility used in the simulations presented in this work. We further acknowledge the project ``Indian participation in the ALICE experiment at CERN,’’ funded by the Department of Science and Technology (DST), Government of India, for financial support under Sanction Order No.~3015/1/2021/Gen/R\&D-I/13283. The authors also acknowledge the Council of Scientific and Industrial Research (CSIR), Government of India, for financial support through a research fellowship to two authors. \end{acknowledgments}
\nocite{*}

\bibliography{paper}% Produces the bibliography via BibTeX.

@article{Shuryak:1978ij,
    author = "Shuryak, Edward V.",
    title = "{Quark-Gluon Plasma and Hadronic Production of Leptons, Photons and Psions}",
    reportNumber = "IYF-78-24",
    doi = "10.1016/0370-2693(78)90370-2",
    journal = "Phys. Lett. B",
    volume = "78",
    pages = "150",
    year = "1978"
}

@article{Barber:1979yr,
    author = "Barber, D. P. and others",
    title = "{Discovery of Three Jet Events and a Test of Quantum Chromodynamics at PETRA Energies}",
    reportNumber = "MIT-LNS-106",
    doi = "10.1103/PhysRevLett.43.830",
    journal = "Phys. Rev. Lett.",
    volume = "43",
    pages = "830",
    year = "1979"
}

@article{Niida:2021wut,
    author = "Niida, T. and Miake, Y.",
    title = "{Signatures of QGP at RHIC and the LHC}",
    eprint = "2104.11406",
    archivePrefix = "arXiv",
    primaryClass = "nucl-ex",
    doi = "10.1007/s43673-021-00014-3",
    journal = "AAPPS Bull.",
    volume = "31",
    number = "1",
    pages = "12",
    year = "2021"
}

@article{PHENIX:2003pfh,
    author = "Adler, S. S. and others",
    collaboration = "PHENIX",
    title = "{J / psi production in Au Au collisions at s(NN)**(1/2) = 200-GeV at the Relativistic Heavy Ion Collider}",
    eprint = "nucl-ex/0305030",
    archivePrefix = "arXiv",
    doi = "10.1103/PhysRevC.69.014901",
    journal = "Phys. Rev. C",
    volume = "69",
    pages = "014901",
    year = "2004"
}

@article{Baym:1999up,
    author = "Baym, Gordon and Heiselberg, Henning",
    title = "{Event-by-event fluctuations in ultrarelativistic heavy ion collisions}",
    eprint = "nucl-th/9905022",
    archivePrefix = "arXiv",
    doi = "10.1016/S0370-2693(99)01263-0",
    journal = "Phys. Lett. B",
    volume = "469",
    pages = "7--11",
    year = "1999"
}

@article{Koch:2001zn,
    author = "Koch, V. and Bleicher, M. and Jeon, S.",
    editor = "Karsch, F. and Satz, H.",
    title = "{Event-by-event fluctuations and the QGP}",
    eprint = "nucl-th/0103084",
    archivePrefix = "arXiv",
    doi = "10.1016/S0375-9474(02)00716-9",
    journal = "Nucl. Phys. A",
    volume = "698",
    pages = "261--268",
    year = "2002"
}

@article{Woithe:2017lzd,
    author = "Woithe, Julia and Wiener, Gerfried J. and Van der Veken, Frederik F.",
    title = "{Let{\textquoteright}s have a coffee with the Standard Model of particle physics!}",
    doi = "10.1088/1361-6552/aa5b25",
    journal = "Phys. Educ.",
    volume = "52",
    number = "3",
    pages = "034001",
    year = "2017"
}

@article{Gross:1973ju,
    author = "Gross, D. J. and Wilczek, Frank",
    title = "{Asymptotically Free Gauge Theories - I}",
    reportNumber = "NAL-PUB-73-49-THY, FERMILAB-PUB-73-049-T",
    doi = "10.1103/PhysRevD.8.3633",
    journal = "Phys. Rev. D",
    volume = "8",
    pages = "3633--3652",
    year = "1973"
}

@article{Gross:1973id,
    author = "Gross, David J. and Wilczek, Frank",
    editor = "Taylor, J. C.",
    title = "{Ultraviolet Behavior of Nonabelian Gauge Theories}",
    doi = "10.1103/PhysRevLett.30.1343",
    journal = "Phys. Rev. Lett.",
    volume = "30",
    pages = "1343--1346",
    year = "1973"
}

@article{Gross:1973zrg,
    author = "Gross, D. J. and Wilczek, Frank",
    title = "{Asymptotically Free Gauge Theories~2}",
    doi = "10.1103/PhysRevD.9.980",
    journal = "Phys. Rev. D",
    volume = "9",
    pages = "980--993",
    year = "1974"
}

@book{Weinberg:1977ji,
    author = "Weinberg, Steven",
    title = "{The First Three Minutes. A Modern View of the Origin of the Universe}",
    journal = " ,,,",
    publisher = "Basic Books",
    isbn = "978-0-465-02437-7",
    year = "1977"
}

@article{Wan:2025rzg,
    author = "Wan, Jie and Wang, Chun-Zheng and Shou, Qi-Ye and Ma, Yu-Gang",
    title = "{Tracing pT-differential radial flow from blast-wave analytics to quark coalescence}",
    eprint = "2509.24889",
    archivePrefix = "arXiv",
    primaryClass = "nucl-th",
    doi = "10.1103/hk6w-hx6h",
    journal = "Phys. Rev. C",
    volume = "113",
    number = "6",
    pages = "064902",
    year = "2026"
}

@article{Odyniec:2019kfh,
    author = "Odyniec, Grazyna",
    editor = "Anagnostopoulos, Konstantinos and others",
    collaboration = "STAR",
    title = "{Beam Energy Scan Program at RHIC (BES I and BES II) {\textendash} Probing QCD Phase Diagram with Heavy-Ion Collisions}",
    doi = "10.22323/1.347.0151",
    journal = "PoS",
    volume = "CORFU2018",
    pages = "151",
    year = "2019"
}

@article{Bialas:1985jb,
    author = "Bialas, A. and Peschanski, Robert B.",
    title = "{Moments of Rapidity Distributions as a Measure of Short Range Fluctuations in High-Energy Collisions}",
    reportNumber = "SACLAY-SPH-T-85-101",
    doi = "10.1016/0550-3213(86)90386-X",
    journal = "Nucl. Phys. B",
    volume = "273",
    pages = "703--718",
    year = "1986"
}

@article{Hwa:2016khr,
    author = "Hwa, Rudolph C. and Yang, C. B.",
    title = "{Observable Properties of Quark-Hadron Phase Transition at the Large Hadron Collider}",
    eprint = "1601.04671",
    archivePrefix = "arXiv",
    primaryClass = "nucl-th",
    doi = "10.5506/APhysPolB.48.23",
    journal = "Acta Phys. Polon. B",
    volume = "48",
    pages = "23",
    year = "2017"
}

@article{Bialas:1988wc,
    author = "Bialas, A. and Peschanski, Robert B.",
    title = "{Intermittency in Multiparticle Production at High-Energy}",
    reportNumber = "SACLAY-SPH-T-88-33, TPJU-4-88",
    doi = "10.1016/0550-3213(88)90131-9",
    journal = "Nucl. Phys. B",
    volume = "308",
    pages = "857--867",
    year = "1988"
}

@article{DeWolf:1995nyp,
    author = "De Wolf, E. A. and Dremin, I. M. and Kittel, W.",
    title = "{Scaling laws for density correlations and fluctuations in multiparticle dynamics}",
    eprint = "hep-ph/9508325",
    archivePrefix = "arXiv",
    reportNumber = "HEN-362A, IIHE-93-01, FIAN-TD-09-93, HEN-362",
    doi = "10.1016/0370-1573(95)00069-0",
    journal = "Phys. Rept.",
    volume = "270",
    pages = "1--141",
    year = "1996"
}

@article{Antoniou:2006zb,
    author = "Antoniou, N. G. and Diakonos, F. K. and Kapoyannis, A. S. and Kousouris, K. S.",
    title = "{Critical opalescence in baryonic QCD matter}",
    eprint = "hep-ph/0602051",
    archivePrefix = "arXiv",
    doi = "10.1103/PhysRevLett.97.032002",
    journal = "Phys. Rev. Lett.",
    volume = "97",
    pages = "032002",
    year = "2006"
}

@article{Antoniou:2005am,
    author = "Antoniou, N. G. and Contoyiannis, Y. F. and Diakonos, F. K. and Mavromanolakis, G.",
    title = "{Critical QCD in nuclear collisions}",
    eprint = "hep-ph/0505185",
    archivePrefix = "arXiv",
    reportNumber = "UA-NPPS-03-2005",
    doi = "10.1016/j.nuclphysa.2005.07.003",
    journal = "Nucl. Phys. A",
    volume = "761",
    pages = "149--161",
    year = "2005"
}

@article{Wu:2022aio,
    author = "Wu, Jin and Li, Zhiming and Luo, Xiaofeng and Xu, Mingmei and Wu, Yuanfang",
    title = "{Intermittency of charged particles in the hybrid UrQMD+CMC model at energies available at the BNL Relativistic Heavy Ion Collider}",
    eprint = "2209.07135",
    archivePrefix = "arXiv",
    primaryClass = "nucl-th",
    doi = "10.1103/PhysRevC.106.054905",
    journal = "Phys. Rev. C",
    volume = "106",
    number = "5",
    pages = "054905",
    year = "2022"
}

@article{Podlaski:2024kxg,
    author = "Podlaski, Piotr",
    title = "{NA61/SHINE Overview}",
    eprint = "2402.10973",
    archivePrefix = "arXiv",
    primaryClass = "nucl-ex",
    doi = "10.1051/epjconf/202429601008",
    journal = "EPJ Web Conf.",
    volume = "296",
    pages = "01008",
    year = "2024"
}

@article{ReynaOrtiz:2024hul,
    author = "Reyna Ortiz, V. Z.",
    collaboration = "NA61/SHINE",
    title = "{Search for the Critical Point via Intermittency Analysis in NA61/SHINE}",
    doi = "10.15407/ujpe69.11.858",
    journal = "Ukr. J. Phys.",
    volume = "69",
    number = "11",
    pages = "858",
    year = "2024"
}

@article{STAR:2023jpm,
    author = "Abdulhamid, Muhammad and others",
    collaboration = "STAR",
    title = "{Energy dependence of intermittency for charged hadrons in Au+Au collisions at RHIC}",
    eprint = "2301.11062",
    archivePrefix = "arXiv",
    primaryClass = "nucl-ex",
    doi = "10.1016/j.physletb.2023.138165",
    journal = "Phys. Lett. B",
    volume = "845",
    pages = "138165",
    year = "2023"
}

@article{Wu2024,
  author = {Wu, J. and Luo, X. and Li, Z. and Wu, Y.},
  title = {[Article title]},
  journal = {Nuclear Physics Review},
  volume = {41},
  number = {4},
  pages = {580},
  year = {2024}
}

@article{NA61SHINE:2024xdd,
    author = "Adhikary, H. and others",
    collaboration = "NA61/SHINE",
    title = "{Search for a critical point of strongly-interacting matter in central \(^{40}\)Ar + \(^{45}\)Sc collisions at 13 A{\textendash}75 A GeV/c beam momentum}",
    eprint = "2401.03445",
    archivePrefix = "arXiv",
    primaryClass = "nucl-ex",
    reportNumber = "FERMILAB-PUB-24-0021-AD",
    doi = "10.1140/epjc/s10052-024-13012-0",
    journal = "Eur. Phys. J. C",
    volume = "84",
    number = "7",
    pages = "741",
    year = "2024"
}

@article{Sharma:2023ndr,
    author = "Sharma, Sheetal and Gupta, Ramni",
    collaboration = "ALICE",
    title = "{Local Multiplicity Fluctuations in~Pb{\ensuremath{-}}Pb Collisions at~$\sqrt{s_\mathrm{{NN}}}$ = 2.76 TeV with~ALICE at~the~LHC}",
    eprint = "2307.14407",
    archivePrefix = "arXiv",
    primaryClass = "nucl-ex",
    doi = "10.1007/978-981-97-0289-3_221",
    journal = "Springer Proc. Phys.",
    volume = "304",
    pages = "860--862",
    year = "2024"
}

@article{Hwa:1992uq,
    author = "Hwa, Rudolph C. and Nazirov, M. T.",
    title = "{Intermittency in second order phase transition}",
    reportNumber = "OITS-490",
    doi = "10.1103/PhysRevLett.69.741",
    journal = "Phys. Rev. Lett.",
    volume = "69",
    pages = "741--744",
    year = "1992"
}

@article{McLerran:1993ka,
    author = "McLerran, Larry D. and Venugopalan, Raju",
    title = "{Gluon distribution functions for very large nuclei at small transverse momentum}",
    eprint = "hep-ph/9311205",
    archivePrefix = "arXiv",
    reportNumber = "TPI-MINN-93-52-T, NUC-MINN-93-28-T, UMN-TH-1224-93",
    doi = "10.1103/PhysRevD.49.3352",
    journal = "Phys. Rev. D",
    volume = "49",
    pages = "3352--3355",
    year = "1994"
}

@article{Werner:2023zvo,
    author = "Werner, Klaus",
    title = "{Revealing a deep connection between factorization and saturation: New insight into modeling high-energy proton-proton and nucleus-nucleus scattering in the EPOS4 framework}",
    eprint = "2301.12517",
    archivePrefix = "arXiv",
    primaryClass = "hep-ph",
    doi = "10.1103/PhysRevC.108.064903",
    journal = "Phys. Rev. C",
    volume = "108",
    number = "6",
    pages = "064903",
    year = "2023"
}

@article{Bass:1998ca,
    author = "Bass, S. A. and others",
    title = "{Microscopic models for ultrarelativistic heavy ion collisions}",
    eprint = "nucl-th/9803035",
    archivePrefix = "arXiv",
    doi = "10.1016/S0146-6410(98)00058-1",
    journal = "Prog. Part. Nucl. Phys.",
    volume = "41",
    pages = "255--369",
    year = "1998"
}

@misc{Werner:2024fwk,
      title={EPOS4: New theoretical concepts for modeling proton-proton and ion-ion scattering at very high energies}, 
      author={Klaus Werner},
      year={2024},
      eprint={2410.09955},
      archivePrefix={arXiv},
      primaryClass={hep-ph},
      url={https://arxiv.org/abs/2410.09955}, 
}

@article{Werner:2023mod,
    author = "Werner, K.",
    title = "{Parallel scattering, saturation, and generalized Abramovskii-Gribov-Kancheli (AGK) theorem in the EPOS4 framework, with applications for heavy-ion collisions at sNN of 5.02 TeV and 200 GeV}",
    eprint = "2310.09380",
    archivePrefix = "arXiv",
    primaryClass = "hep-ph",
    doi = "10.1103/PhysRevC.109.034918",
    journal = "Phys. Rev. C",
    volume = "109",
    number = "3",
    pages = "034918",
    year = "2024"
}

@article{Hwa:2011bu,
    author = "Hwa, Rudolph C. and Yang, C. B.",
    title = "{Local Multiplicity Fluctuations as a Signature of Critical Hadronization at LHC}",
    eprint = "1111.6651",
    archivePrefix = "arXiv",
    primaryClass = "nucl-th",
    doi = "10.1103/PhysRevC.85.044914",
    journal = "Phys. Rev. C",
    volume = "85",
    pages = "044914",
    year = "2012"
}

@article{NA49:2012ebu,
    author = "Anticic, T. and others",
    collaboration = "NA49",
    title = "{Critical fluctuations of the proton density in A+A collisions at 158$A$ GeV}",
    eprint = "1208.5292",
    archivePrefix = "arXiv",
    primaryClass = "nucl-ex",
    doi = "10.1140/epjc/s10052-015-3738-5",
    journal = "Eur. Phys. J. C",
    volume = "75",
    number = "12",
    pages = "587",
    year = "2015"
}

@article{Leykam:2022ejk,
    author = "Leykam, Daniel and Angelakis, Dimitris G.",
    title = "{Topological data analysis and machine learning}",
    eprint = "2206.15075",
    archivePrefix = "arXiv",
    primaryClass = "cond-mat.mes-hall",
    doi = "10.1080/23746149.2023.2202331",
    journal = "Adv. Phys. X",
    volume = "8",
    number = "1",
    pages = "2202331",
    year = "2023"
}

@article{Hamilton:2022blu,
    author = "Hamilton, Greg and Dore, Travis and Plumberg, Christopher",
    title = "{Applications of persistent homology in nuclear collisions}",
    eprint = "2209.15480",
    archivePrefix = "arXiv",
    primaryClass = "nucl-th",
    doi = "10.1103/PhysRevC.106.064912",
    journal = "Phys. Rev. C",
    volume = "106",
    number = "6",
    pages = "064912",
    year = "2022"
}

@article{Capellino:2025kce,
    author = "Capellino, Federica and Dubla, Andrea and Masciocchi, Silvia and Nijs, Govert and Spitz, Daniel",
    title = "{Toward a topological data analysis for heavy-ion collisions}",
    eprint = "2509.02339",
    archivePrefix = "arXiv",
    primaryClass = "nucl-th",
    reportNumber = "CERN-TH-2025-168",
    doi = "10.1103/w5pw-prkv",
    journal = "Phys. Rev. C",
    volume = "112",
    number = "5",
    pages = "054909",
    year = "2025"
}

@article{Wang:2024bzy,
    author = "Wang, Rui and Qiu, Chengrui and Hu, Chuan-Shen and Li, Zhiming and Wu, Yuanfang",
    title = "{Identifying weak critical fluctuations of intermittency in heavy-ion collisions with topological machine learning}",
    eprint = "2412.06151",
    archivePrefix = "arXiv",
    primaryClass = "nucl-th",
    doi = "10.1016/j.physletb.2025.139405",
    journal = "Phys. Lett. B",
    volume = "864",
    pages = "139405",
    year = "2025"
}

@misc{Duy:2016,
      title={Limit theorems for persistence diagrams}, 
      author={Trinh Khanh Duy and Yasuaki Hiraoka and Tomoyuki Shirai},
      year={2016},
      eprint={1612.08371},
      archivePrefix={arXiv},
      primaryClass={math.PR},
      url={https://arxiv.org/abs/1612.08371}, 
}

@inproceedings{Chen:2016btl,
author = {Chen, Tianqi and Guestrin, Carlos},
title = {XGBoost: A Scalable Tree Boosting System},
year = {2016},
isbn = {9781450342322},
publisher = {Association for Computing Machinery},
address = {New York, NY, USA},
url = {https://doi.org/10.1145/2939672.2939785},
doi = {10.1145/2939672.2939785},
booktitle = {Proceedings of the 22nd ACM SIGKDD International Conference on Knowledge Discovery and Data Mining},
pages = {785–794},
numpages = {10},
location = {San Francisco, California, USA},
series = {KDD '16}
}

@misc{lundberg:2017,
      title={A Unified Approach to Interpreting Model Predictions}, 
      author={Scott Lundberg and Su-In Lee},
      year={2017},
      eprint={1705.07874},
      archivePrefix={arXiv},
      primaryClass={cs.AI},
      url={https://arxiv.org/abs/1705.07874}, 
}

@article{Antoniou:1998np,
    author = "Antoniou, N. G. and Contoyiannis, Y. F. and Diakonos, F. K. and Papadopoulos, C. G.",
    title = "{Fractals at T = T(c) due to instanton - like configurations}",
    eprint = "hep-ph/9810383",
    archivePrefix = "arXiv",
    doi = "10.1103/PhysRevLett.81.4289",
    journal = "Phys. Rev. Lett.",
    volume = "81",
    pages = "4289--4292",
    year = "1998"
}

@article{Antoniou:2000ms,
    author = "Antoniou, N. G. and Contoyiannis, Y. F. and Diakonos, F. K. and Karanikas, A. I. and Ktorides, C. N.",
    title = "{Pion production from a critical QCD phase}",
    eprint = "hep-ph/0012164",
    archivePrefix = "arXiv",
    reportNumber = "UA-NPPS-13-2000",
    doi = "10.1016/S0375-9474(01)00921-6",
    journal = "Nucl. Phys. A",
    volume = "693",
    pages = "799--824",
    year = "2001"
}

@article{Sharma:2023oxo,
    author = "Sharma, Sheetal and Malik, Salman Khurshid and Banoo, Zarina and Gupta, Ramni",
    title = "{Normalized factorial moments of spatial distributions of particles in high multiplicity events: A Toy model study}",
    eprint = "2309.07712",
    archivePrefix = "arXiv",
    primaryClass = "hep-ph",
    doi = "10.1016/j.nuclphysa.2024.122963",
    journal = "Nucl. Phys. A",
    volume = "1053",
    pages = "122963",
    year = "2025"
}

@article{Malik:2024ltm,
    author = "Malik, Salman Khurshid and Sharma, Sheetal and Gupta, Ramni",
    collaboration = "ALICE",
    title = "{Event-by-event local multiplicity fluctuations in charged particle production at the LHC energies with ALICE}",
    journal = "DAE Symp. Nucl. Phys.",
    volume = "67",
    pages = "1003--1004",
    year = "2024"
}

@article{Wu:2021jou,
    author = "Wu, Jin and Lin, Yufu and Li, Zhiming and Luo, Xiaofeng and Wu, Yuanfang",
    title = "{Intermittency analysis of proton numbers in heavy-ion collisions at energies available at the BNL Relativistic Heavy Ion Collider}",
    eprint = "2104.11524",
    archivePrefix = "arXiv",
    primaryClass = "nucl-th",
    doi = "10.1103/PhysRevC.104.034902",
    journal = "Phys. Rev. C",
    volume = "104",
    number = "3",
    pages = "034902",
    year = "2021"
}

@article{Sharma:2018vtf,
    author = "Sharma, Rohni and Gupta, Ramni",
    title = "{Scaling Properties of Multiplicity Fluctuations in the AMPT Model}",
    eprint = "1806.10854",
    archivePrefix = "arXiv",
    primaryClass = "hep-ph",
    doi = "10.1155/2018/6283801",
    journal = "Adv. High Energy Phys.",
    volume = "2018",
    pages = "6283801",
    year = "2018"
}

@article{Gupta:2019zox,
    author = "Gupta, Ramni and Malik, Salman Khurshid",
    title = "{Intermittency study of charged particles generated in Pb-Pb collisions at $\sqrt{s_{\mathrm{NN}}}\text{= 2.76 TeV}$ using EPOS3}",
    eprint = "1911.13111",
    archivePrefix = "arXiv",
    primaryClass = "hep-ex",
    doi = "10.1155/2020/5073042",
    journal = "Adv. High Energy Phys.",
    volume = "2020",
    pages = "5073042",
    year = "2020"
}

@article{Haider:2026rtw,
    author = "Haider, Fakhar Ul and Malik, Salman Khurshid and Gupta, Ramni and Singh, Balwan",
    title = "{Study of the charged particle multiplicity fluctutions in EPOS4 for Pb-Pb $sqrt{s_{NN}}$ = 5.02 TeV}",
    journal = "DAE Symp. Nucl. Phys.",
    volume = "69",
    pages = "1081--1082",
    year = "2026"
}

@article{Sarma:2019teo,
    author = "Sarma, Pranjal and Bhattacharjee, Buddhadeb",
    title = "{Color reconnection as a possible mechanism of intermittency in the emission spectra of charged particles in PYTHIA-generated high-multiplicity $pp$ collisions at energies available at the CERN Large Hadron Collider}",
    eprint = "1902.09124",
    archivePrefix = "arXiv",
    primaryClass = "hep-ph",
    doi = "10.1103/PhysRevC.99.034901",
    journal = "Phys. Rev. C",
    volume = "99",
    number = "3",
    pages = "034901",
    year = "2019"
}

@article{Singh:2024gai,
    author = "Singh, Arpit and Kumar, Ashwini and Chandra, Anuj and Singh, Sweta and Ahmad, Shakeel and Singh, B. K.",
    title = "{Scaling properties of particle density fluctuations at LHC energies}",
    doi = "10.1209/0295-5075/ad7757",
    journal = "EPL",
    volume = "148",
    number = "1",
    pages = "14001",
    year = "2024"
}

@article{ALICE:2013axi,
    author = "Abbas, E. and others",
    collaboration = "ALICE",
    title = "{Performance of the ALICE VZERO system}",
    eprint = "1306.3130",
    archivePrefix = "arXiv",
    primaryClass = "nucl-ex",
    reportNumber = "CERN-PH-EP-2013-082",
    doi = "10.1088/1748-0221/8/10/P10016",
    journal = "JINST",
    volume = "8",
    pages = "P10016",
    year = "2013"
}

@article{ALICE:2018tvk,
    author = " ",
    collaboration = "ALICE",
    title = "{Centrality determination in heavy ion collisions}",
    reportNumber = "ALICE-PUBLIC-2018-011, ALICE-PUBLIC-2018-011",
    journal = " ",
    month = "8",
    year = "2018"
}

@article{Huang:2021iux,
    author = "Huang, Yige and Pang, Long-Gang and Luo, Xiaofeng and Wang, Xin-Nian",
    title = "{Probing criticality with deep learning in relativistic heavy-ion collisions}",
    eprint = "2107.11828",
    archivePrefix = "arXiv",
    primaryClass = "nucl-th",
    doi = "10.1016/j.physletb.2022.137001",
    journal = "Phys. Lett. B",
    volume = "827",
    pages = "137001",
    year = "2022"
}

@article{ALICE:2014sbx,
    author = "Abelev, Betty Bezverkhny and others",
    collaboration = "ALICE",
    title = "{Performance of the ALICE Experiment at the CERN LHC}",
    eprint = "1402.4476",
    archivePrefix = "arXiv",
    primaryClass = "nucl-ex",
    reportNumber = "CERN-PH-EP-2014-031",
    doi = "10.1142/S0217751X14300440",
    journal = "Int. J. Mod. Phys. A",
    volume = "29",
    pages = "1430044",
    year = "2014"
}

@article{Lippmann:2014lay,
    author = "Lippmann, Christian",
    collaboration = "ALICE",
    title = "{Upgrade of the ALICE Time Projection Chamber}",
    reportNumber = "CERN-LHCC-2013-020, ALICE-TDR-016",
    journal = " ",
    month = "3",
    year = "2014"
}

@article{Bialas:1990dk,
    author = "Bialas, A. and Gazdzicki, M.",
    title = "{A New variable to study intermittency}",
    reportNumber = "CERN-TH-5859-90",
    doi = "10.1016/0370-2693(90)90575-Q",
    journal = "Phys. Lett. B",
    volume = "252",
    pages = "483--486",
    year = "1990"
}

@article{Cheng:2006qk,
    author = "Cheng, M. and others",
    title = "{The Transition temperature in QCD}",
    eprint = "hep-lat/0608013",
    archivePrefix = "arXiv",
    reportNumber = "BNL-NT-06-27, BI-TP-2006-31, CU-TP-1158",
    doi = "10.1103/PhysRevD.74.054507",
    journal = "Phys. Rev. D",
    volume = "74",
    pages = "054507",
    year = "2006"
}

@article{Gribov:1999ui,
    author = "Gribov, V. N.",
    editor = "Nyiri, J.",
    title = "{The Theory of quark confinement}",
    eprint = "hep-ph/9902279",
    archivePrefix = "arXiv",
    reportNumber = "BONN-TK-98-09",
    doi = "10.1007/s100529900052",
    journal = "Eur. Phys. J. C",
    volume = "10",
    pages = "91--105",
    year = "1999"
}

@article{Koch:2025cog,
    author = "Koch, Volker and Vovchenko, Volodymyr",
    title = "{Exploring the QCD phase diagram through correlations and fluctuations}",
    eprint = "2512.04288",
    archivePrefix = "arXiv",
    primaryClass = "nucl-th",
    doi = "10.1140/epjs/s11734-026-02307-w",
    journal = "Eur. Phys. J. Spec. Top.",
    month = "12",
    year = "2025"
}

\end{document}